\documentclass[11pt]{article}

\usepackage{arxiv}

\makeatletter
\def\blfootnote{\gdef\@thefnmark{}\@footnotetext}
\makeatother

\usepackage[dvipsnames,  x11names, table]{xcolor}
\usepackage{amsmath,amsfonts}
\usepackage{algorithmic}
\usepackage{algorithm}
\usepackage{array}
\usepackage[caption=false,font=normalsize,labelfont=sf,textfont=sf]{subfig}
\usepackage{textcomp}
\usepackage{stfloats}
\usepackage{url}
\usepackage{xurl}
\usepackage{verbatim}
\usepackage{graphicx}
\usepackage{cite}
\usepackage{float}
\usepackage{enumitem}
\usepackage{flushend}
\usepackage{booktabs}
\usepackage{makecell}

\newcolumntype{L}[1]{>{\raggedright\let\newline\\\arraybackslash\hspace{0pt}}m{#1}}
\newcolumntype{C}[1]{>{\centering\let\newline\\\arraybackslash\hspace{0pt}}m{#1}}
\newcolumntype{R}[1]{>{\raggedleft\let\newline\\\arraybackslash\hspace{0pt}}m{#1}} 

\newtheorem{definition}{Definition}

\begin{document}

\title{A Framework for Responsible AI Systems: Building Societal Trust through Domain Definition, Trustworthy AI Design, Auditability, Accountability, and Governance}

\newcommand{\shorttitle}{A Framework for Responsible AI Systems}

\author{Andr\'es Herrera-Poyatos, Javier Del Ser, Marcos {López de Prado},\\  \textbf{Fei-Yue Wang, Enrique Herrera-Viedma, and~Francisco~Herrera}}

\blfootnote{This publication is part of the IAFER TSI-100927-2023-1 project, funded by the Recovery, Transformation, and Resilience Plan of the Next Generation of the EU through the Spanish Ministry of Digital Transformation and the Civil Service (F. Herrera, E. Herrera-Viedma); and supported by the Science and Technology Development Fund of Macau SAR (No. 0145/2023/RIA3) (F-Y. Wang) and the Basque Government (MATHMODE, IT1866-26) (J. Del Ser).}
\blfootnote{Corresponding author: F. Herrera.}
\blfootnote{Andr\'es Herrera-Poyatos, Enrique Herrera-Viedma and Francisco Herrera are with the Deparment of Computer Science and Artificial Intelligence and Andalusian Research Institute in Data Science and Computational Intelligence (DaSCI), University of Granada, 18071 Granada, Spain. E-mail: \{andreshp,viedma,herrera\}@ugr.es.}
\blfootnote{Javier Del Ser is with TECNALIA (BRTA), 48160 Bizkaia, Spain; and with the University of the Basque Country (UPV/EHU), 48940 Leioa, Spain. E-mail: javier.delser@tecnalia.com.} 
\blfootnote{Marcos {López de Prado} is with the School of Engineering, Cornell University, Ithaca, NY, 14850, United States; and ADIA Lab, Al Maryah Island, Abu Dhabi, United Arab Emirates; and the Computational Research Department, Lawrence Berkeley National Laboratory, Berkeley, CA 94720, United States. E-mail: ml863@cornell.edu.}
\blfootnote{Fei-Yue Wang is with the Intelligent Systems for Robotics and Automation Laboratory, Macau University of Science and Technology, Macau SAR; and The State Key Laboratory for Management and Control of Complex Systems, CASIA, Beijing, China. E-mail: feiyue.wang@ia.ac.cn.}

\maketitle

\begin{abstract}
Responsible Artificial Intelligence (RAI) addresses the ethical and regulatory challenges of deploying AI systems in high-risk scenarios. This paper proposes a comprehensive framework for the design of an RAI system (RAIS) that integrates five key dimensions: domain definition, trustworthy AI design, auditability, accountability, and governance. Unlike prior work that treats these components in isolation, our proposal emphasizes their inter-dependencies and iterative feedback loops, enabling proactive and reactive accountability throughout the AI lifecycle. Beyond presenting the framework, we synthesize recent developments in global AI governance and analyze limitations in existing principles-based approaches, highlighting fragmentation, implementation gaps, and the need for participatory governance. The paper also identifies critical challenges and research directions for the RAIS framework, including sector-specific adaptation and operationalization, to support certification, post-deployment monitoring, and risk-based auditing. By bridging technical design and institutional responsibility, this work offers a practical blueprint for embedding responsibility throughout the AI lifecycle, enabling transparent, ethically aligned, and legally compliant AI-based systems.
\end{abstract}


\section{Introduction}

Artificial Intelligence (AI) has evolved into a mature and sophisticated technology, quietly entered our lives, and made a great leap in the last year. Generative AI models have shown that AI has gone, in just a few months, practically from science fiction to becoming an essential part of our daily lives.

This emergence goes hand in hand with a growing global debate on the ethical dimension of AI. Concerns arise about its impact on data privacy, fundamental rights, and protection against discrimination in automated decisions, or the continued presence of fake videos and images. The risks of AI are relatively well known, such as the potential for automated decisions to be harmful to certain vulnerable groups, the hidden biases that can arise from the data used in its training, or the security due to the vulnerability of AI systems to adversarial attacks \cite{critch2023tasra,hendrycks2023overview, hendrycks2025introduction, bengio2024managing}. This scenario raises the need for responsible, fair, inclusive, trusted, safe and secure, transparent, and accountable frameworks. 

In recent years, the concept of Trustworthy AI (TAI) \cite{ai2019high} has emerged as a framework to ensure the ethical, safe and accountable development, deployment, and use of AI systems. TAI is built on three core pillars (lawfulness, ethics, and robustness) and encompasses requirements such as human oversight, technical robustness and safety, privacy and data governance, transparency, fairness, societal well-being, and accountability.

The increasing use of AI in high-risk domains underscores the need for frameworks that ensure that AI systems are not only trustworthy by design but also responsibly applied. In this regard, the term responsible AI (RAI) \cite{dignum2019responsible} is often mistakenly treated as synonymous with TAI. RAI emphasizes the ethical and legal use of AI systems, focusing on auditability and accountability, while TAI represents a broader paradigm that integrates legal aspects with technical requirements. Translating responsibility into AI-based systems requires that recommendations derived from system output remain accountable, legally compliant, and ethical.

A key concept guiding RAI and TAI is the definition of an RAI system (RAIS), as introduced in \cite{diaz2023connecting}: 
\begin{definition}
A \textit{responsible AI system} is an AI-based system that ensures auditability and accountability during its design, development, and use, according to specifications and the applicable regulation of the domain of practice in which the AI system is to be used.
\end{definition}

As highlighted by Novelly et al. (2024) \cite{novelli2024accountability}, accountability is a multifaceted concept that involves an answerability relationship between an agent and a forum, which requires recognition of authority, interrogation, and limitation of power. It also encompasses features such as context, standards, processes, and implications that shape how responsibility is exercised and enforced. 
In this work, we adopt a liability-based perspective as a pragmatic choice for regulatory alignment, while acknowledging that a comprehensive approach should embed mechanisms for justification, oversight, and ethical scrutiny.

Building on this foundation, our central hypothesis is that responsible AI in high-risk scenarios cannot be achieved through isolated principles or technical tools; it demands an integrated RAIS framework that aligns trustworthiness, auditability, governance, stakeholder context, and domain constraints across the AI lifecycle.

To address this, we systematically explore these dimensions and propose a holistic RAIS framework comprising five key components: i) domain scenario; ii) TAI based design along with standardization and assessments; iii) auditability and certification; iv) accountability; and v) AI governance, requiring a transversal vision across all stages of development and deployment. This perspective supports the need for an adaptive and resilient RAIS framework that can evolve with societal demands, legal norms, and technological transformations.

Our discussions are driven by three research questions:

\begin{itemize}
    \item[(i)] How can auditability and accountability be operationalized across the AI lifecycle? 
    \item[(ii)]  What methodologies can effectively integrate technical trustworthiness with socio-legal requirements? 
    \item[(iii)] How can dynamic feedback loops and participatory governance be embedded to ensure continuous compliance and enable societal trust in AI?
\end{itemize}

By examining these questions, our paper pursues three objectives:
\begin{itemize}[leftmargin=*]
\item First, it serves as a position paper that proposes a comprehensive RAIS framework to enable societal trust in AI, introducing the five interconnected  key dimensions. This framework highlights that an RAIS can gain trust and ethics AI \cite{choung2023trust},  enable societal trust in AI systems.  

\item Second, the paper provides a detailed analysis of the five fundamental dimensions that support and guide the development and deployment of an RAIS. Each dimension is discussed both as a core component of the proposed framework and in terms of its specific challenges.

\item Third, we complement our analysis with a synthesis of design insights, reflections and the challenges to be addressed to advance the development of an RAIS.

\end{itemize}

The organization of this paper is as follows. Section~\ref{sec:RAIS} introduces the basic questions on the concept of responsible AI systems, paying attention to  (\textit{why?}), and (\textit{how?}). Section \ref{sec:RAIS-F} proposes the RAIS framework to enable societal trust in AI. Sections \ref{sec:context}, \ref{sec:TAI}, \ref{sec:AUD}, \ref{sec:ACC} and \ref{sec:AIg} elaborates on each of the key dimensions of the proposed RAIS framework, respectively: domain definition, TAI design, paying attention to AI safety and the essential role of explanations, auditability and certification, accountability and inspection, and AI governance. Section \ref{sec:example} exemplifies the application of the framework in the context of autonomous driving, whereas Section \ref{sec:Key} summarizes the design insights, framework reflections, and challenges that must be addressed to enable the development of an RAIS. Section \ref{sec:Con} ends the paper with a short conclusion and an outlook on this emerging area of enormous practical relevance.

\section{Responsible AI Systems: Why and How?}
\label{sec:RAIS}

In this section, we discuss the concept of an RAIS in a double perspective. First, Subsection \ref{sec:RAIS-HRS} tackles the need for responsibility, why?. Second, Subsection \ref{sec:RAIS-sota} analyzes the studies proposed in the literature to design and implement the RAIS framework, how?

\subsection{Why do we need Responsible AI Systems? Enabling Societal Trust in AI, Regulation and High-Risk Scenarios}\label{sec:RAIS-HRS}

Legal and regulatory frameworks are essential in the context of responsible AI. These frameworks provide the necessary guidelines and standards to ensure that AI technologies are developed and deployed in a manner that is compliant with existing laws and regulations. Effective legal and regulatory measures can help build social trust in AI technologies and encourage their responsible use.

In recent progress, intense advances and debates have been held on AI regulation by governments and institutions. Among others, the European Commission AI Act\footnote{\url{https://artificialintelligenceact.eu/} [acc. 30/12/25]} \cite{AIA24}, the China AI Regulation with the ``Code of Ethics for New-generation Artificial Intelligence''\footnote{\url{https://www.most.gov.cn/kjbgz/202109/t20210926_177063.html} [acc. 30/12/25]}, the UAE's International Stance on Artificial Intelligence Policy\footnote{\url{https://uaelegislation.gov.ae/en/policy/details/uae-s-international-stance-on-artificial-intelligence-policy} [acc. 30/12/25]}, the U.S. Department of Treasury\textit{"Framework to advance AI governance and risk management in National Security"}\footnote{\url{https://ai.gov/wp-content/uploads/2024/10/NSM-Framework-to-Advance-AI-Governance-and-Risk-Management-in-National-Security.pdf} [acc. 30/12/25]}. or the AI Principles of the OECD\footnote{\url{https://www.oecd.org/en/topics/ai-principles.html} [acc. 30/12/25]}. In fact, the OECD AI Principles were updated in 2024, to include 1) inclusive growth, sustainable development, and well-being; 2) human rights and democratic values, fairness, and privacy; 3) transparency and explainability; 4) robustness, security, and safety; and 5) accountability; among others. There is an increasing demand for AI regulations to minimize risks, ensure safety, and preserve human rights while fostering a flexible and innovative environment. These efforts must align with risk analysis \cite{critch2023tasra,hendrycks2023overview} to create a global framework for the secure implementation and deployment of AI systems in all applications where risks may arise.

\begin{figure}[H]
	\centering
    \vspace{-2mm}
	\includegraphics[width=0.7\columnwidth]{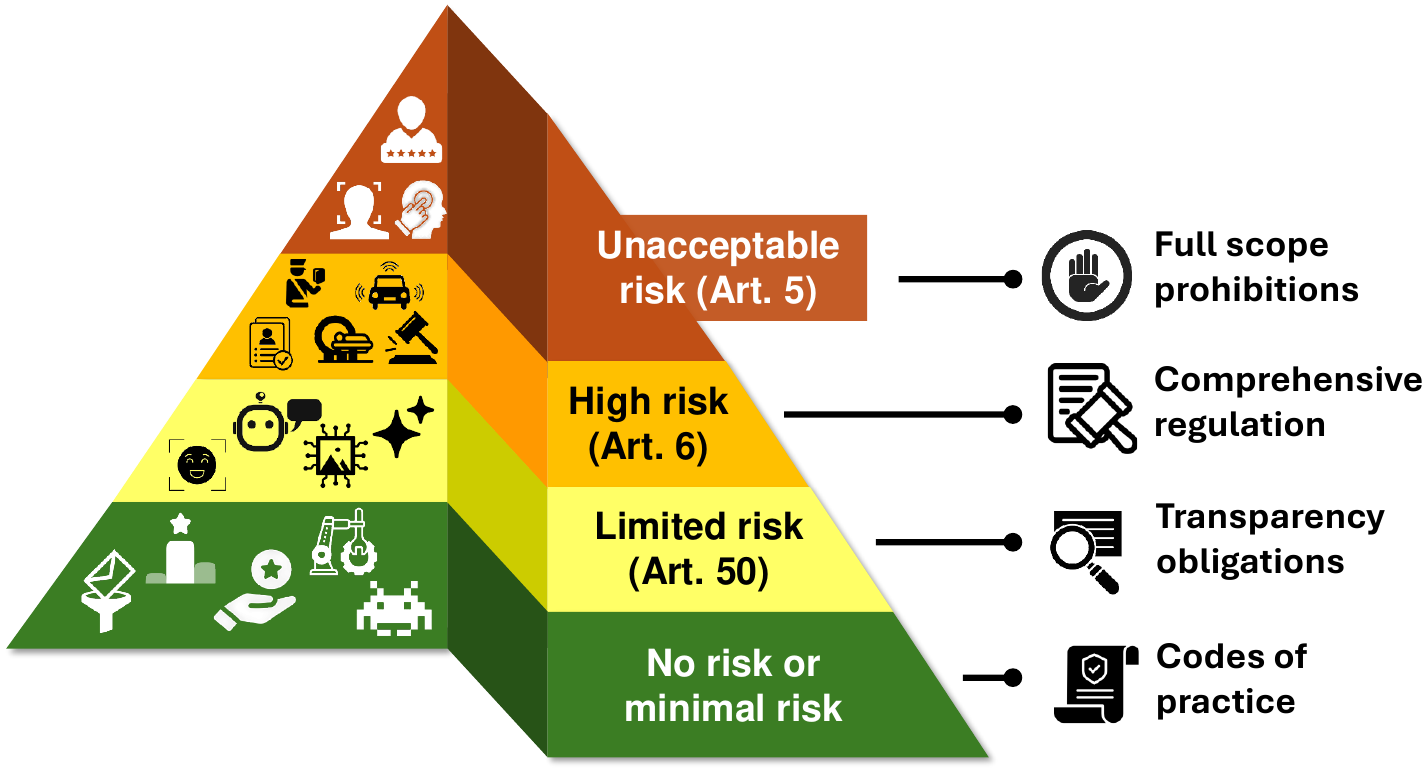}
	\caption{Risk levels defined by the AI Act regulatory framework.}
	\label{fig:risks}
\end{figure}

The European AI Act is one of the most advanced regulations worldwide, setting comprehensive standards for the development, deployment, and use of AI \cite{AIA24}. It was published in June 2024, as the first attempt to enact a horizontal AI regulation. The proposed legal framework establishes a technology neutral definition of AI systems in EU legislation and defines a classification for AI systems based on risk level problems (Figure \ref{fig:risks}). We embrace this regulatory RAIS framework as a basis  in high-risk scenarios and to analyze their auditability and accountability prerequisites. 

As shown in Figure \ref{fig:risks}, the European AI Act classifies systems into four risk levels, with obligations proportional to the level of risk. 

\begin{itemize}[leftmargin=*]
\item \emph{Levels 3 and 4}: Most AI applications, such as recommendation engines or spam filters, pose minimal risk and generally face no mandatory obligations, though voluntary codes of conduct may apply.
\item \emph{Level 2}: High-risk systems require audits and adherence to horizontal and sector-specific regulations. Examples include biometric identification, critical infrastructure, education and employment systems, essential services, law enforcement, migration control, and judicial processes. These systems can significantly affect safety or fundamental rights (Art. 6).
\item \emph{Level 1}: Systems presenting unacceptable risk (such as government social scoring or toys encouraging dangerous behavior) are prohibited.
\end{itemize}

Applying general AI systems to high-risk scenarios can bring about significant benefits, but also poses substantial risks if such systems are deployed without adequate safeguards. In these contexts, failures can directly impact fundamental rights, safety, and social welfare, making trust a prerequisite rather than a byproduct of adoption. Societal trust in AI is the degree to which people (as individuals and as a public) believe that AI systems are safe, fair, competent, and governed in ways that align with their interests; it is  a multidimensional construct operating across individual, institutional, and experiential levels \cite{choung2023trust}. Consequently, enabling societal trust in high-risk AI systems requires the systematic integration of RAI principles into both technical design and regulatory compliance.

\subsection{How has Responsible AI been approached to date? }
\label{sec:RAIS-sota}

The RAI research area has evolved rapidly in recent years, driven by growing societal awareness of AI's ethical implications and its deployment in high-risk domains. Governments, industry, and civil society organizations have proposed numerous guidelines, exemplified by the OECD Principles.  However, different studies have highlighted that these abstract and often overlapped principles lack operational specificity and have had a limited impact on AI development practices in organizations \cite{sadek2024challenges,goellner2024responsible}.

To address this translation gap, several requirements assessment lists and applied frameworks  have been proposed. The ALTAI (Assessment List for Trustworthy Artificial Intelligence) \cite{radclyffe2023assessment} and Z-Inspection \cite{zicari2021z} frameworks operationalize TAI requirements \cite{ai2019high}, providing assessment tools in domains such as healthcare and public services. The AI Risk Management Framework (AIRMF) released by the American National Institute of Standards and Technology (NIST) \cite{ai2023artificial} emphasizes principles or organizational readiness. 
The ECCOLA framework aligns agile development practices with ethical reflection \cite{vakkuri2021eccola}. The MATCH framework  evaluates organizational maturity for responsible AI \cite{liao2022designing}. 

However, these tools often operate in isolation, are inconsistently adopted, and lack integration into standard engineering pipelines, thereby reinforcing a fragmented landscape. Furthermore, the principle-based approach has come under increasing scrutiny for its limited
practical utility. Sadek et al. (2024) \cite{sadek2024challenges} identify five systemic challenges that inhibit the implementation of AI in practice: i) the abstract nature of guidelines, ii) conflicting and narrow value definitions, iii) absence of actionable metrics, iv) fragmentation of responsibilities throughout the AI pipeline and v) lack of internal advocacy and accountability. The review points to the need for participatory and sociotechnical methods that align the design of the AI system with the values of stakeholders and practical constraints. These findings
corroborate earlier criticisms that RAI practices often devolve into performative compliance -- relying on checklists and codes of ethics -- without achieving meaningful ethical alignment \cite{mittelstadt2019principles,morley2023operationalising}.

In response to these implementation challenges, the authors have increasingly focused on the operational and organizational dimensions of RAI.  Lu et al. (2023) \cite{lu2023responsible} introduce a "Responsible-AI-by-design" pattern collection aimed at software engineers and system architects. Their patterns provide actionable guidance for embedding responsibility during system development, offering a bridge between ethical theory and technical design, a bridge also pursued by frameworks like ECCOLA and MATCH, but with a more direct application to development practices. Brey and Dainow (2024) \cite{brey2024ethics} proposed integrating ethical checkpoints throughout the AI development lifecycle .  Woodgate and Ajmeri (2024) \cite{woodgate2024macro} expand the ethical discourse by categorizing macro-level ethical principles and advocating for their alignment with sociotechnical realities. They argue that an RAIS cannot be sustained through checklists or compliance audits alone; instead, systems must be evaluated through stakeholder-sensitive. reflective mechanisms capable of adapting to evolving contexts and values. Papagiannidis et al. (2025) \cite{papagiannidis2025responsible} propose a multi-layered framework that emphasizes structural, procedural, and relational governance mechanisms, highlighting the gap between ethical principles and their organizational embodiment. They suggest that governance must be embedded within the workflows, cultures, and responsibilities of institutions to move beyond aspirational compliance.

In this context, it is important to recognize that part of the apparent ``fragmentation'' of the field is not solely attributable to methodological deficits, but also to the intrinsically sociotechnical nature of the problem. Dai and Xiao (2025) \cite{dai2025embracing} argue that theoretical inconsistency should not be interpreted as a weakness of the RAI movement, but rather as evidence of normative pluralism and of inevitable tensions between values (e.g., transparency vs. privacy, fairness vs. performance, control vs. autonomy) that arise in complex systems. This perspective shifts the objective from ``unifying principles'' to explicitly managing trade-offs and contradictions through operational mechanisms such as contextual definition, verifiable requirements, auditing, responsibility allocation, and governance with feedback loops. For this reason, in the next section we propose an integrative RAIS framework that articulates these dimensions and their interdependencies across the system life cycle, in order to enable compliance, accountability, and societal trust in high-risk scenarios.

\begin{figure}[H]
	\centering
	\includegraphics[width=\columnwidth]{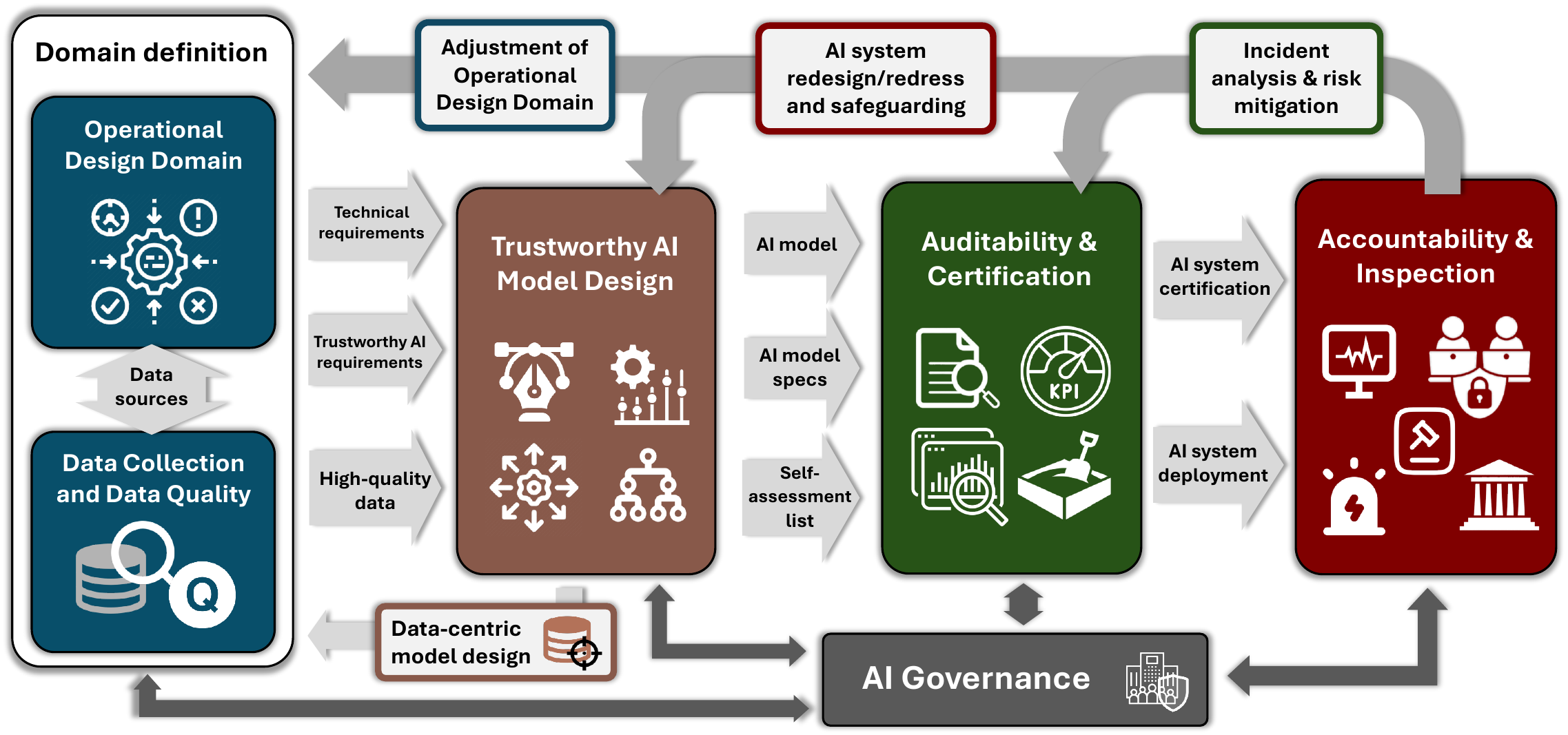}
	\caption{Proposed RAIS framework.}
	\label{fig:framework}
\end{figure}

\section{Responsible AI System Framework: The Consensus for Societal Trust in AI}
\label{sec:RAIS-F}

We now proceed by introducing the proposed RAIS framework in  Subsection~\ref{sec:RAIS-frame}.  We briefly compare the RAIS framework with previous discussed frameworks in Subsection~\ref{sec:contrast});  ending this section by  highlighting their crucial role in promoting societal trust in AI in Subsection~\ref{sec:Societal}. 

\subsection{Responsible AI Systems Framework}
\label{sec:RAIS-frame}

The concept of the RAIS framework must be analyzed from a holistic perspective that integrates five key dimensions: domain definition, trustworthy AI design, auditability, accountability, and governance. Together, these dimensions provide the foundation for the development and deployment of an RAIS, enabling adaptation to diverse risk scenarios and application contexts.

Figure \ref{fig:framework} illustrates this framework, which organizes the five dimensions and their analytical components into four interconnected layers, and the transversal AI governance supervising and coordinating all them in a feedback loop: 

\begin{itemize}[leftmargin=*]
    \item Domain layer. It defines the problem in which an AI system operates, including its intended purpose, stakeholders, operational constraints, data quality, and risk profile. 
    \item Trustworthy AI design layer: Focuses on technical requirements, high-quality data, and compliance with TAI principles during model development.
    \item Auditability \& certification layer: Ensures conformity through self-assessment lists, documentation, and adherence to applicable standards and norms. 
    \item Accountability \& inspection layer: Covers system certification, deployment, and continuous monitoring, supported by mechanisms for incident analysis, risk mitigation, and iterative improvement.
\end{itemize}

We highlight four important aspects of the design of the proposed framework:

\begin{itemize}[leftmargin=*]
     
\item \textit{AI governance as a cross-cutting dimension:} AI governance is positioned as a foundational dimension that influences and receives feedback from all other components of the framework. Provides the institutional and procedural backbone to define responsibilities, enforce compliance, and monitor system behavior throughout the AI lifecycle.

\item \textit{Feedback and Iteration Mechanisms:} The framework incorporates dynamic feedback loops that connect the accountability stage to earlier phases of the system lifecycle. When issues are detected, such as ethical breaches, performance failures, or context misalignment, these mechanisms trigger targeted responses that include: (i) incident analysis and risk mitigation procedures, (ii) AI system redesign or redress interventions, and (iii) updates to the operational design domain (ODD) and data quality parameters. 

\textit{AI Safety Considerations:} These feedback and iteration mechanisms are vital for advancing AI safety, particularly in open-world and high-risk environments. As discussed in more detail in Section \ref{sec:RAIS-Safety}, AI safety goes beyond traditional system performance to include the prevention of unintended behaviors, the containment of emerging risks, and the assurance of system reliability under real-world conditions. 

\item \textit{Data-Centric Model Design:} It is also essential to consider a data-centric approach to model development, where the focus shifts from tweaking model architectures to improving the quality, representativeness, and ethical integrity of the data itself. This paradigm emphasizes the curating of datasets that are diverse, well-labeled, and aligned with the intended application context.

\item Other elements discussed throughout the next sections are also shown in the framework (e.g., the use of self-assessment lists in the model design phase, or certification and safety as primary goals of auditability and accountability). Among them, XAI is a central element in the framework. XAI acts as a bridge between technical decision-making and human oversight, allowing developers, users, and regulators to understand the rationale behind the AI output. This is particularly critical in high-risk or dynamic scenarios, where explainability directly impacts trust, error detection, and ethical compliance. Moreover, XAI is a fundamental enabler of Human-in-the-Loop (HITL) mechanisms, as discussed later in Subsection \ref{sec:RAIS-XAI}.

\end{itemize}

The specific relationships among the dimensions and elements of the framework must be developed in detail for each application in each high-risk scenario. This is due to the necessary nuances with respect to specific regulation, the other technical aspects required, from data quality and ODD to the model design, and benchmarking. We advocate embedding stakeholder engagement and domain risk analysis throughout the AI lifecycle. Similarly, the framework can be adapted to other specific regulatory and governance requirements with a similar flow of steps. It must be aligned  with international standards such as ISO/IEC DIS 42001 \cite{ISO42001-IA} and IEEE 7000 \cite{IEEE-Std-7000-2021}. The framework is conceived as a valuable tool for guiding the systematic design and implementation of an RAIS for each potential application, ensuring that all critical aspects are addressed.

\subsection{Comparison to other RAIS frameworks}
\label{sec:contrast}

Existing approaches emphasize principles or organizational readiness and often treat these dimensions in isolation. In contrast, our framework integrates technical and governance aspects and embeds dynamic feedback loops for proactive and reactive accountability.

Table~\ref{tab:framework_comparison} highlights these differences, showing how our proposal advances previous work through explicit compliance mapping, structured auditability pathways, and continuous improvement mechanisms.

\begin{table}[ht!]
\vspace{-2mm}
\centering
\setlength{\tabcolsep}{1pt}
\caption{Comparison of our framework with existing approaches.}
\label{tab:framework_comparison}
\begin{tabular}{C{1.5cm}C{2cm}C{2cm}C{2cm}C{2cm}C{2cm}}
\toprule
\textbf{Framework} & \textbf{Lifecycle Integration} & \textbf{Compliance Mapping} & \textbf{Multi-Stakeholder Involvement} & \textbf{Auditability Pathways} & \textbf{Dynamic Feedback Mechanisms} \\
\midrule
\makecell[c]{\textbf{ALTAI}\\\cite{radclyffe2023assessment}} & Primarily design and early deployment & EU AI Act principles & Limited; mainly developer self-assessment & High-level self-assessment, no formal certification & Minimal; static checklist \\
\midrule
\makecell[c]{\textbf{NIST AI}\\\textbf{RMF} \cite{ai2023artificial}}  & Broad; risk management across lifecycle & U.S. regulatory alignment, voluntary & Encourages stakeholder engagement & Risk management practices; no sector-specific audit workflows & Iterative risk assessment, but not deeply embedded \\
\midrule
\makecell[c]{\textbf{ECCOLA}\\\cite{vakkuri2021eccola}} & Integrated into agile development & Ethics-by-design principles & Strong; participatory design focus & Informal ethical checkpoints & Feedback loops within agile sprints \\
\midrule
\makecell[c]{\textbf{MATCH}\\\cite{liao2022designing}} & Organizational maturity focus, not technical lifecycle & Governance and compliance readiness & Emphasizes organizational roles & No detailed technical audit workflow & Feedback primarily via periodic maturity reassessments \\
\midrule
\textbf{RAIS framework} & Full lifecycle; from context to governance & Explicit mapping to regulatory obligations & Embedded; roles for developers, auditors, regulators & Structured auditability leading to certification & Continuous feedback loops for redesign and risk mitigation \\
\bottomrule
\end{tabular}
\end{table}

As a result, our proposed framework responds to the aforementioned research questions by (i) proposing a lifecycle-based architecture for auditability and accountability, (ii) integrating technical and governance dimensions into a unified design, and (iii) embedding iterative feedback and participatory mechanisms to advance societal trust in AI systems. Our framework, tailored for high-risk applications, addresses gaps in current practices by moving beyond static assessments toward dynamic, lifecycle responsibility, yielding a coherent, multi-dimensional architecture that bridges ethical principles with implementable processes. It is a scalable blueprint for embedding responsibility into organizational workflows, regulatory compliance, and technical design, supporting certification, monitoring, and sector-specific adaptation.

\subsection{Societal Trust in AI}
\label{sec:Societal}

Societal trust in AI is a prerequisite for the legitimate deployment of AI systems, particularly in domains that affect high-risk and rights-affecting domains. Rather than arise solely from technical performance, trust reflects collective judgments about whether AI systems are safe, fair, transparent, and governed in ways that align with societal values and expectations \cite{afroogh2024trust}. Empirical evidence shows that trust varies strongly in application contexts and increases when algorithmic decisions are perceived as understandable, contestable, and institutionally controlled \cite{orban2025trust}. At the individual level, trust and appropriate reliance are further shaped by psychological traits and risk perceptions, reinforcing the need for AI systems that support informed human judgment rather than automation bias \cite{kuper2025psychological}.

From a longitudinal perspective, the literature has progressively shifted from abstract notions of trust to operational mechanisms that enable it in practice, such as accountability structures, governance processes, and post-deployment oversight \cite{benk2025twenty}. Crucially, philosophical analyses emphasize that trust must be grounded in genuine trustworthiness: fostering confidence in AI systems without provable assurances risks ethical misalignment and loss of legitimacy \cite{duran2025trust}. Consequently, societal trust should be understood as a rational response to verifiable properties of AI systems and their institutional embedding, rather than as an outcome of persuasive or purely reputational strategies.

In this work, societal trust is therefore treated as an emergent outcome of an RAIS that integrates trustworthy design with enforceable safeguards throughout the AI lifecycle. In particular, auditability mechanisms (Section~\ref{sec:AUD}) provide the technical and procedural means to verify compliance, detect failures, and support accountability, while governance structures (Section~\ref{sec:AIg}) establish the institutional conditions for oversight, stakeholder participation, and continuous alignment with societal values. Together, these dimensions operationalize societal trust as a measurable and enforceable property of AI systems, rather than a purely subjective perception.

\section{Domain Definitions for Responsible AI Systems}
\label{sec:context}

We discuss three fundamental and domain aspects that we need to consider before designing AI systems, the ODD in Subsection \ref{sec:ODD}, data collection quality in Subsection \ref{sec:Dat} and stakeholders context in Subsection \ref{sec:roles}. 
 ODD and data quality are tailored to the specific application domain, shaping and informing the model design with the TAI functionalities to be implemented. In this way, TAI technologies do not operate disconnected from the real problem, domain reality, and the audience intended for the AI system.

\begin{figure}[h!]
	\centering
    \includegraphics[width=0.5\columnwidth]{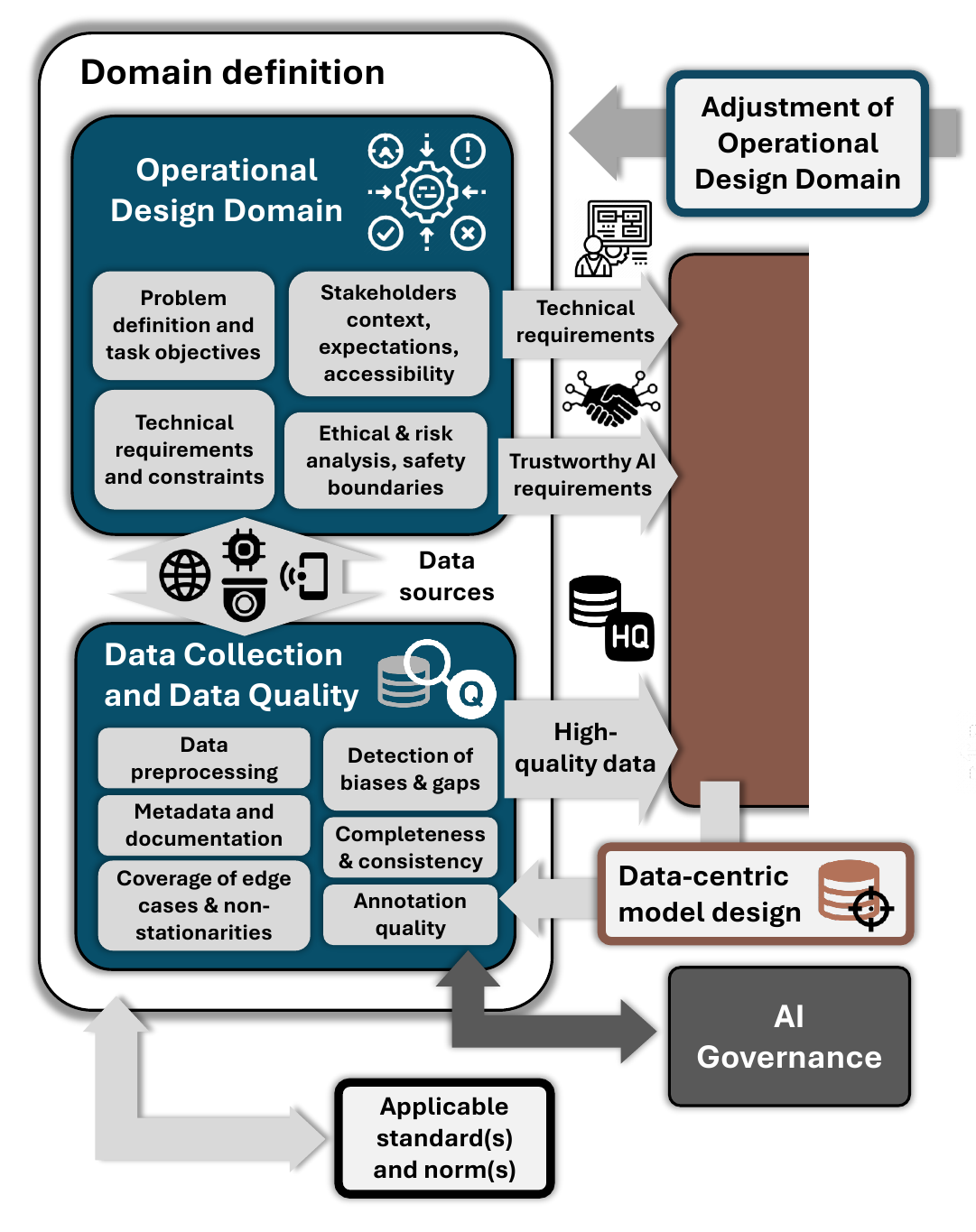}
	\caption{Domain definition and its elements in the proposed RAIS framework.}
	\label{fig:Domain}
    \vspace{-3mm}
\end{figure}

 Figure\ref{fig:Domain} summarizes graphically the contextual elements related to domain definition that are tackled in this section.

\subsection{Operational Design Domain} 
\label{sec:ODD}

One of the two main steps to achieve an RAIS is the definition of ODD. It refers to the domain and specific conditions under which an AI system is designed to operate safely and effectively. Widely used in autonomous driving \cite{betz2024new}, ODD includes factors such as data sources, users, functional and non-functional requirements, and the specific task(s) that the AI system is expected to tackle. Delineating the ODD of an AI system is crucial to ensure that the system is used within their intended limits and can perform reliably and safely.

Therefore, understanding the domain in which an AI system operates provides critical information on the specific scenarios and conditions that the AI system will encounter, allowing for a more precise and tailored design and a safer and better aligned operation. Taking into account the ODD factors, developers can create an RAIS that are better equipped to handle real-world situations. This domain awareness helps identify potential risks and limitations, ensuring that the AI system can perform optimally and safely within its defined ODD. The range of factors to consider for ODD is wide, depending on the specific problem. Among others:
\begin{itemize}[leftmargin=*]
\item Environment: the physical and digital environments in which the AI system will be deployed and operated.
\item Geographical location: the specific regions or areas where the AI system will be deployed, considering local regulations and cultural contexts.
\item Data sources, including providers, access, and characteristics.
\item User interaction: types of interaction that the AI system will maintain with users, including voice, text, and physical interfaces.
\item Task(s) that the AI system is designed to perform, including their complexity and the functional capabilities of the AI system when addressing such task(s).
\item  Data requirements and management (e.g., if data will be stored or whether the AI model adapts to the arrival of new data over time).
\item Technological constraints: any limitations or constraints on the AI system's operation, such as memory footprint, energy consumption, or inference latency.
\item Regulatory compliance: adherence to local, national and international regulations and standards relevant to the operation of the AI system.
\end{itemize}

Other aspects, such as safety and ethical considerations, are associated with the TAI requirements, which will be examined in detail in Section \ref{sec:TAI}. Due to the application-specific nature of ODD, we can find studies on ODD for specific applications, such as autonomous driving \cite{betz2024new}. 

\subsection{Data Quality}
\label{sec:Dat}

In the context of AI, data quality stands for the accuracy, completeness, consistency, and reliability of the data used to train and operate AI systems. Access to high-quality data is instrumental for the development of effective TAI based systems, as it directly impacts their performance and decision-making capabilities. Ensuring data quality helps reduce biases, improve the accuracy of predictions, and improve the overall reliability of AI systems. By maintaining rigorous standards for data quality, we can build AI systems that are fair, transparent and capable of providing valuable information, while minimizing the risks associated with erroneous or biased data.

 Data quality preprocessing techniques are essential to improve data and provide the perfect context for AI system design \cite{garcia2015data,luengo2020big}. Preprocessing steps such as data cleaning, normalization, and enhancement help to refining the raw data, making it more suitable for AI training. These processes ensure that the data set is free from errors, minimize inconsistencies, and remove irrelevant information, thus enhancing its quality. This proactive approach to data quality management is crucial for building AI systems that are responsible, efficient, and capable of addressing complex real-world problems.
 
Paying attention to the data domain is equally important \cite{serra2024use}. The domain in which data are collected, processed, and used can significantly influence the results of AI models. Understanding the source, relevance, and applicability of data ensures that AI systems are trained on information that is representative of real-world scenarios. This domain awareness helps identify potential biases and gaps in the data, allowing more informed and accurate AI predictions.

When an AI system is trained to make decisions at time $t$ on data that were not available at time $t$, its estimated forecast power will be inflated. This creates the danger of developing and using AI systems that will systematically underperform and potentially cause harm, hence having severe safety implications. Therefore, an RAIS that makes decisions at a time $t$ must be trained only on data that was available at that time $t$. This concept is known as \emph{point-in-time}  data \cite{de2018advances}, and plays a pivotal role in applications where the underlying probability distributions to be modeled change significantly over time, such as financial time series or industrial machinery data for predictive maintenance.  

In relation to \emph{task complexity}, an RAIS should be able to recognize when the system is asked to extrapolate, that is, when the system is presented with a set of input values that are significantly underrepresented or not found in the training set (out-of-distribution inputs). Under those circumstances, an RAIS should warn the user that it is making a prediction for which it has little statistical support, or decline to produce such prediction. Thus, progress on the field of out-of-distribution learning is fundamental in the domain of an RAIS, see, e.g., \cite{barcina2024managing,hendrycks2021many}. 

Continuing the same line of thought, an RAIS should incorporate mechanisms to detect when a structural breakdown has occurred, making its training data untrustworthy. This is  important in the context of highly dynamic and non-stationary data-generating processes, such as finance and economy \cite{lopez2023causal}.

\subsection{Stakeholders context}
\label{sec:roles}

In \cite{haresamudram2023three}, a stakeholder interest map is proposed to distinguish six levels of audience profile: developer, designer, owner, user, regulator, and society.  This distinction is not merely taxonomic, but reflects fundamentally different relationships with AI systems in terms of agency, responsibility, and exposure to risk. Understanding why these distinctions matter is essential for designing Responsible AI systems that are interpretable, auditable, and accountable across contexts.

This stakeholder stratification raises a set of critical questions:
\textit{How should AI systems adapt their behavior and explanations across audiences?}
In particular:
\begin{itemize}[leftmargin=*]
  \item Are all stakeholders equal in terms of their required level of understanding?
  \item What constitutes a stakeholder’s ``understood'' requirement?
  \item What qualifies as an appropriate ``explanation'' for each audience?
  \item What aspects of the system’s behavior are observable to different stakeholders?
\end{itemize}
 
This need to contextualize understanding is reinforced by Deshpande and Sharp (2022) \cite{deshpande2022responsible}, who argue that Responsible AI systems operate within a multi-stakeholder ecosystem spanning individual, organizational, and national or international actors. Their analysis highlights that stakeholders differ not only in influence and exposure, but also in informational needs, interpretative capacities, and responsibilities throughout the AI lifecycle. Consequently, what constitutes a sufficient explanation, observable behavior, or adequate level of understanding varies substantially between developers, users, regulators, and society at large.

Following audience-aware XAI \cite{bello2025three}, we map stakeholder roles (audience profiles) to \emph{primary} layered explanation needs (algorithmic/domain, human-centered, and social), while preserving traceability across layers to support role-sensitive transparency and accountable governance. As discussed in \cite{bello2025three}, explicitly accounting for stakeholder context is fundamental to adapting representations of system behavior to different audiences in governance-relevant ways, and to avoiding one-size-fits-all explanations that undermine auditability, accountability allocation, and governance-relevant transparency.

\vspace{2mm}
\noindent \textbf{\textit{Dimension Frontier.}} The interaction between data quality and ODD must be explored in depth for high-risk scenarios, as it directly shapes how different stakeholders specify requirements, assess risks, and evaluate system behavior. For example, the work introduced in \cite{cappi2024design} addresses the critical aspects of ODD and data quality in the context of a vision-based landing task, presenting a robust methodology for designing and validating datasets. This framework addresses the challenges of designing a dataset compliant with the requirements of AI system certification in safety-critical applications, thus supporting developers, certifiers, and regulators in establishing verifiable assurance claims.

The importance of ODD specification and data quality assurance has become increasingly critical in the pursuit of high-quality of an RAIS. As emphasized by Stettinger et al. (2024) \cite{stettinger2024trustworthiness}, the assurance of trustworthiness in high-risk AI applications is highly dependent on robust definitions of ODD that capture environmental, operational, and functional boundaries, ensuring predictable and verifiable system behavior. Such definitions play a central role in aligning system capabilities with stakeholder expectations, particularly for operators, affected users, and oversight bodies. Meanwhile, Wang et al. (2024) \cite{wang2024overview} provide a comprehensive synthesis of data quality dimensions (including accuracy, completeness, consistency, timeliness, and relevance) which directly impact the reliability and interpretability of AI-driven decisions. These dimensions are foundational for both upstream model development and downstream auditing, enabling different stakeholder groups to reason about system limitations and accountability.

However, significant challenges remain, including the integration of user-centered quality assessments, the management of heterogeneous data sources, and the guaranty of the domain suitability of data across deployment contexts. Together, these insights reinforce the need for a structured and multidisciplinary approach to data governance and ODD formulation as cornerstones of a reliable and ethically aligned AI deployment. Importantly, such an approach must explicitly account for stakeholder roles and perspectives, establishing the conditions required to design, assess, and adapt an RAIS in a manner that is appropriate to each domain of application and transparent to those who develop, use, regulate and are affected by AI technologies.

\section{Trustworthy AI Design}
\label{sec:TAI}

TAI involves creating systems that are reliable, safe and secure, operating transparently and accountably. This requires a comprehensive approach that includes rigorous testing, continuous monitoring, and adherence to best practices and standards.By prioritizing trustworthiness in AI development, we can ensure that these technologies are used in ways that benefit society while minimizing risks and potential harms for high-risk scenarios.

We introduce the TAI requirements considered to design an RAIS. Then, we pay attention to two fundamental requirements,  explainability, and safety. In Subsection~\ref{sec:RAIS-XAI}, we discuss the role of XAI as a fundamental enabler of HITL decision making in an RAIS. The Subsection~\ref{sec:RAIS-Safety} then emphasizes AI safety as a primary goal in an RAIS.

\subsection{Trustworthy AI Paradigm}
\label{sec:TAI-Req}

\begin{figure}[htb!]
	\centering
    \includegraphics[width=0.7\columnwidth]{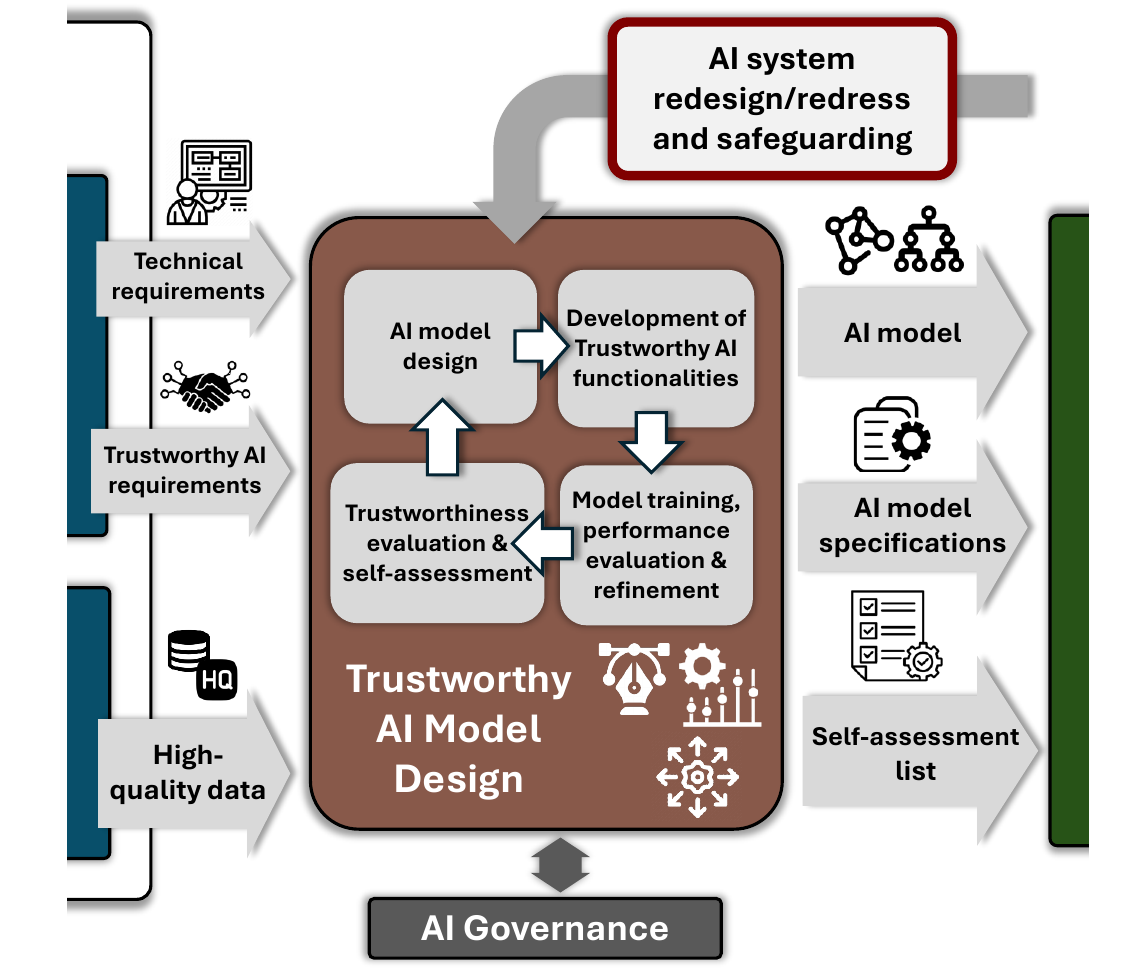}
	\caption{Trustworthy AI model design within the proposed framework.}
	\label{fig:trustAI}
\end{figure}

We introduce the TAI paradigm and its technical requirements  \cite{diaz2023connecting}. Figure \ref{fig:trustAI} graphically shows this dimension within the proposed framework.

We begin with a precise definition of TAI:
\vspace{2mm}
\begin{definition} \label{def:tai}
    \textit{Trustworthy AI} is a paradigm that encompasses all technical approaches and tools to develop, deploy, and use safe, legal, and ethical AI systems.
\end{definition}
\vspace{2mm}

Going deeper into the 3 fundamental pillars of TAI \cite{ai2019high}, AI-based systems must adhere to robust, legal, and ethical technical standards, ensuring that:
 \begin{enumerate}[leftmargin=*]
\item  AI-based systems are legal and comply with all applicable laws and regulations (\emph{lawfulness}).
\item  AI-based systems are ethical, ensuring the adherence to ethical principles and values (\emph{ethics}).
\item  AI-based systems are robust and safe, from both technical and social perspectives (\emph{robustness}).
 \end{enumerate}

Each of these three components is necessary, but not sufficient, to achieve TAI. %\subsection{Trustworthy AI Requirements} 
%\label{ssec:reqTAI}
These three pillars of TAI can be broken down into seven technical requirements according to the high-level expert group on the AI proposal \cite{ai2019high}:

\begin{enumerate}[leftmargin=*]
   \item \emph{Human agency and oversight}: AI systems should support human agency and autonomy, and human oversight. 
   \item \emph{Technical robustness and safety}:  General safety, accuracy, reliability, response to attack and security, fallback plans, and reproducibility. Cybersecurity is a critical component of technical robustness in high-risk AI systems and has been part of this requirement since its initial formulation \cite{ai2019high}.

   \item \emph{Privacy and data governance}: AI systems should protect user privacy and handle data responsibly and should adhere to data governance principles. In this third requirement, cybersecurity is also essential, as privacy-preserving mechanisms are needed to ensure data confidentiality, integrity, and controlled access throughout the AI lifecycle.
   \item \emph{Transparency}: Traceability, explainability, communication.
   \item \emph{Diversity, non-discrimination, and fairness}: Avoidance of unfair bias, accessibility and universal design, and stakeholder participation.
   \item \emph{Societal and environmental wellbeing}: AI systems should be designed considering their wider impact of AI on society, democracy, and the environment by actively engaging stakeholders and assessing their implications for work, skills, and sustainability.
   \item \emph{Accountability}: AI systems should be auditable and subject to effective risk management processes, with mechanisms in place to assign responsibility for failures.
\end{enumerate}

Several institutions around the world have promoted different frameworks for risk management in AI systems, in which TAI tools and requirements play a central role. Among them, in January 2023 NIST released AIRMF \cite{ai2023artificial} which, among other TAI requirements, highlights  "secure and resilient", "explainable and interpretable", and "valid and reliable" AI systems.

It is important to note that "secure and resilient" should be treated as an independent requirement, distinct from safety. AI systems, as well as the ecosystems in which they are deployed, may be said to be resilient if they can withstand unexpected adverse events or unexpected changes in their environment or use,  or if they can maintain their functions and structure in the face of internal and external change and degrade safely and gracefully when this is necessary (adapted from: ISO/IEC TS 5723:2022). Common security concerns relate to adversarial examples, data poisoning, and the exfiltration of models, training data, or other intellectual property through AI system endpoints. AI systems that can maintain confidentiality, integrity, and availability through protection mechanisms that prevent unauthorized access and use can be said to be secure. The guidelines in the NIST Cybersecurity Framework and the Risk Management Framework are among those which are applicable here.

Meeting TAI requirements in practice poses significant challenges. Their integration is fundamental for the development of RAIS, particularly when moving from high-level principles to actionable and context-sensitive implementations. Beyond these technical limitations, Wirz et al. (2025) \cite{wirz2025re} argue that trustworthiness cannot be treated as a purely objective property engineered into an AI system, but must instead be understood as perceptual and context-dependent, varying across stakeholders. Consequently, TAI design within RAIS must extend beyond compliance with technical requirements to incorporate mechanisms for stakeholder-aware evaluation, auditability, and adaptive governance across the AI lifecycle.

\subsection{On the Essential Role of Explanations in Responsible AI}
\label{sec:RAIS-XAI}

Explanations play a central role in enabling auditability, accountability, and meaningful human oversight. Within the RAIS framework, explanations provide access to the reasoning processes or mechanisms through which AI system outputs are generated, allowing stakeholders to interrogate system behavior beyond surface-level performance. As such, explainability is a critical capability of Responsible AI, ensuring that AI-driven decisions and actions remain understandable, contestable, and governable by humans.

A widely adopted definition of explainability emphasizes two fundamental elements: \emph{understanding} and \emph{audience}. As formulated by Arrieta et al.~\cite{arrieta2020explainable}:

\vspace{2mm}
\begin{definition}
    Given an audience, an \textit{explainable AI} is one that produces details or reasons to make its functioning clear or easy to understand.
\end{definition}
\vspace{2mm}

This audience-centered perspective is particularly relevant for RAIS, where explanations must serve diverse stakeholders—including developers, auditors, regulators, and domain experts—each with distinct information needs and responsibilities.

In this broader vision, XAI acts as a bridge between complex AI models, human understanding, and human--AI interaction~\cite{kim2023help}. By providing appropriate explanations, AI systems can demonstrate compliance with ethical principles and regulatory requirements, support the identification of biases or errors, and facilitate informed decision-making. Explainability therefore contributes directly to governance processes by enabling scrutiny, oversight, and accountability, rather than simply improving user comprehension.

Although these benefits motivate the widespread adoption of XAI, they do not alone resolve all challenges associated with explainability in practice. At the same time, the rapid expansion of XAI research has raised important questions about its maturity, limitations, and risks. Recent work has examined the current state of XAI~\cite{ali2023explainable,munoz2025maturity}, identified open challenges~\cite{longo2024explainable}, and highlighted potential misuse and unintended consequences~\cite{nannini2025nullius,alpsancar2025explanation}. These contributions converge on the need for a more nuanced and critical deployment of XAI, particularly in high-risk scenarios where explanations are often assumed to guarantee trust or ethical compliance.

The opacity of many AI systems, particularly deep learning models, limits transparency and constrains meaningful user understanding. As a result, achieving the design of TAI within RAIS requires role-sensitive explainability mechanisms and the support of institutional trust structures to ensure responsible use and oversight \cite{herrera2025opacity}.

Empirical results suggest that XAI mainly affects perceived legitimacy rather than guaranteeing trustworthiness: Shulner-Tal et al.~\cite{shulner2025made} find that users tend to rate AI decisions as less fair and less understandable than human decisions, but that suitable explanation styles can substantially narrow this perception gap. This supports role- and context-sensitive explainability in RAIS, while underscoring that explanations must be governed and evaluated against explicit oversight goals.

These limitations have prompted a growing body of critical literature that questions prevailing assumptions about the role and value of explainability. Alpsancar et al.~\cite{alpsancar2025explanation} argue that the value of explanations is fundamentally instrumental and purpose-dependent: explanations matter only relative to specific goals such as accountability, governance, or situated decision-making, which must be made explicit rather than assumed. Similarly, Nannini et al.~\cite{nannini2025nullius} caution that XAI can introduce new technical and sociotechnical risks, including robustness vulnerabilities, circular reasoning, essentialism, and accountability failures. From this perspective, explanations should not be treated as trust guarantees, but as objects of ethical and technical risk management that require continuous monitoring, documentation, and assignment of responsibility.

These concerns are compounded by the lack of consensus on what constitutes ``adequate'' explainability. Requirements vary significantly across application domains, stakeholder groups, and regulatory contexts, making one-size-fits-all solutions inappropriate. Moreover, inherent tensions between TAI requirements further complicate explainability design~\cite{moreno2025design}: increasing transparency may conflict with privacy obligations, fairness interventions may reduce accuracy, and robustness measures can increase system complexity. Addressing these trade-offs requires adaptive governance and iterative design strategies rather than static compliance approaches.

In high-risk scenarios, the degree and nature of explainability are particularly consequential. Explanations may range from identifying salient input features to modeling causal dependencies that reveal the mechanisms driving system behavior. Causal explainability, when feasible, offers stronger guarantees of robustness, reproducibility, and accountability by enabling stakeholders to distinguish statistical correlations from genuine causal drivers~\cite{lopez2023causal}. For this reason, high-risk RAIS should prioritize causal or mechanistic explanations whenever domain constraints allow.

Ultimately, XAI is a foundational enabler for auditability and HITL oversight. Effective auditing requires expert interaction with AI systems to test, validate, and contest outputs under various conditions, using explanations as auditable evidence of system behavior (Section~\ref{sec:AUD}).

In parallel, governance frameworks rely on explanations to support responsibility allocation, compliance verification, and institutional oversight (Section~\ref{sec:AIg}). Without meaningful context-aware explanations, collaborative human--AI decision-making is undermined by opacity, overreliance, or distrust. As emphasized in~\cite{herrera2025}, XAI must therefore evolve beyond technical clarity to support attentiveness, contextual sensitivity, and sustained human engagement throughout the AI lifecycle.

\vspace{2mm}

\vspace{2mm}
\noindent \textbf{\textit{XAI Design Principles for RAIS}.}
From the perspective of RAIS, explainability should be (i) audience- and role-sensitive, (ii) purpose-driven and explicitly linked to governance and accountability goals, (iii) robust to misuse and adversarial exploitation, and (iv) integrated into iterative audit and oversight processes rather than treated as a static system feature. Embedding these principles into sector-specific benchmarks and formal assurance frameworks is essential to ensure that explainability contributes meaningfully to auditability, governance, and ultimately societal trust in high-risk AI systems.

\subsection{AI Safety as a Goal}
\label{sec:RAIS-Safety}

Safety is one of the main goals of an RAIS, essential for auditability and accountability. As these systems increasingly impact our society, ensuring their safe deployment is essential to building trust, mitigating risks, and upholding ethical standards. AI safety is an interdisciplinary field that is concerned with preventing accidents, misuse, or other harmful consequences that could result from the use of AI systems. Beyond AI research, safety implies the development of standards and policies enforcing it in the practical context of use \cite{hendrycks2021unsolved, hendrycks2025introduction,bengio2025singapore}. AI safety encompasses:
\begin{itemize}[leftmargin=*]
  \item \emph{Machine ethics}: It is part of AI ethics concerned with adding or ensuring moral behaviors of man-made machines that use AI. It is also called \emph{machine morality}. 
   \item \emph{AI alignment}: It refers to the guide for AI systems to act according to the intended goals, preferences, or ethical principles of humans. An AI system is aligned when it reliably pursues its intended objectives. A misaligned AI system may pursue unintended or conflicting objectives, potentially leading to malfunctions, unintended consequences, or harm.
   \item \emph{Robustness}, which refers to the ability of an AI system to maintain performance against various perturbations and adversarial input throughout the AI lifecycle.  
     
     \item \emph{AI security}, also known as  \emph{external safety} or  \emph{systemic safety}, addresses broader contextual risks in the way AI systems are managed. Cybersecurity and decision-making play a decisive role in whether AI systems fail or are misdirected.
  
   \item \emph{AI monitoring systems}, which include tools designed to continuously observe and evaluate the performance of AI models, ensuring that they operate reliably and adhere to predefined standards. AI monitoring systems must incorporate defenses against adversarial inputs, unauthorized model access, and data exfiltration. Continuous vulnerability assessments and penetration tests are essential to ensure resilience. Cybersecurity safeguards should be embedded throughout the AI lifecycle, from data collection to deployment, to prevent cascading failures by malicious actors.
   
\end{itemize}

In accordance with the AI safety goals, an RAIS will unleash their full potential when trust can be established at each stage of their lifecycle. This requires: 
\begin{itemize}[leftmargin=*]
    \item Identification of training and monitoring metrics to minimize errors, false positives, and biases.
    \item The performance of tests (e.g. bias testing) to produce verifiable results and increase end-user trust. 
    \item Methodologies to determine the validity of tests for an RAIS.
    \item Continuously monitoring after deployment to ensure that the AI model operates in a responsible and unbiased way.
\end{itemize}

When it comes to continuous monitoring, it is straightforward to wonder what happens if a problem is detected in an AI system. In that case, a critical accountability process is triggered, comprising two steps:
\begin{itemize}[leftmargin=*]
\item The first step is to analyze the incident and mitigate the risk. Incident analysis can require redefining the ODD and acquiring more and/or better quality data. 
\item In the second step, the newly designed an RAIS must undergo a new audit process.
 \end{itemize}
 
This iterative process of detection, redesign, and re-auditing is essential to maintain the integrity and trustworthiness of AI systems. It is essential to ensure that they continue to function ethically and effectively in real-world scenarios. This cycle is at the core of our proposed framework, adopting the iterative AI safety assurance process as its main design driver.

%To end with, we draw attention to the mechanism defined in the AI Act for systematic safety assessment of AI-based systems: the so-called \emph{regulatory sandboxes} \cite{truby2022sandbox} for high-risk scenarios. Sandboxes allow for the evaluation of the conformity of the AI-based system with respect to technical specifications, horizontal and vertical regulation, and ethical principles in a controlled and reliable testing environment. The advantage of isolated environments is that they favor the development, testing, and validation of innovative AI systems under the direct supervision and guidance of the competent authorities (Art. 53 of the AI Act). Two main aspects related to sandboxes remain unresolved to date: i) the design of sandboxing guidelines that rapidly and effectively implement algorithmic auditing for high-risk scenarios; and ii) the development of intelligent systems for high-risk scenarios validated through the necessary auditing processes. 

\vspace{2mm}
\noindent \textbf{\textit{Dimension Frontier.}} As highlighted by Kowald et al. in \cite{kowald2024establishing}, although conceptual frameworks for trustworthy AI are increasingly comprehensive (encompassing fairness, accountability, and transparency), the establishment and validation of these principles in practice remain insufficiently operationalized. They argue for the lifecycle-wide integration of evaluation metrics and domain-sensitive benchmarks, emphasizing the need for methodological consistency across contexts.   Recent studies on explainability, robustness in TAI by Chander et al. (2005) \cite{chander2025toward} highlight persistent gaps, including the lack of standardized methodologies, clear evaluation metrics, and operational guidance for real-world deployment. Their findings reinforce the urgency of developing adaptive tools and dynamic assessment frameworks that can assess trustworthiness throughout the AI lifecycle. 

This perspective underscores that trade-offs between performance, interpretability, and safety cannot be resolved universally but must be managed through processes that explicitly account for stakeholder roles, situational context, and governance structures.

\section{Auditability and Certification}
\label{sec:AUD}

 \emph{Auditability} is becoming increasingly important as standards are implemented. In terms of particular audit tools, especially when an RAIS interacts with the user, grading schemes adapted to the use case are needed to validate an intelligent system.

\begin{figure}[h!]
	\centering
    \includegraphics[width=0.7\columnwidth]{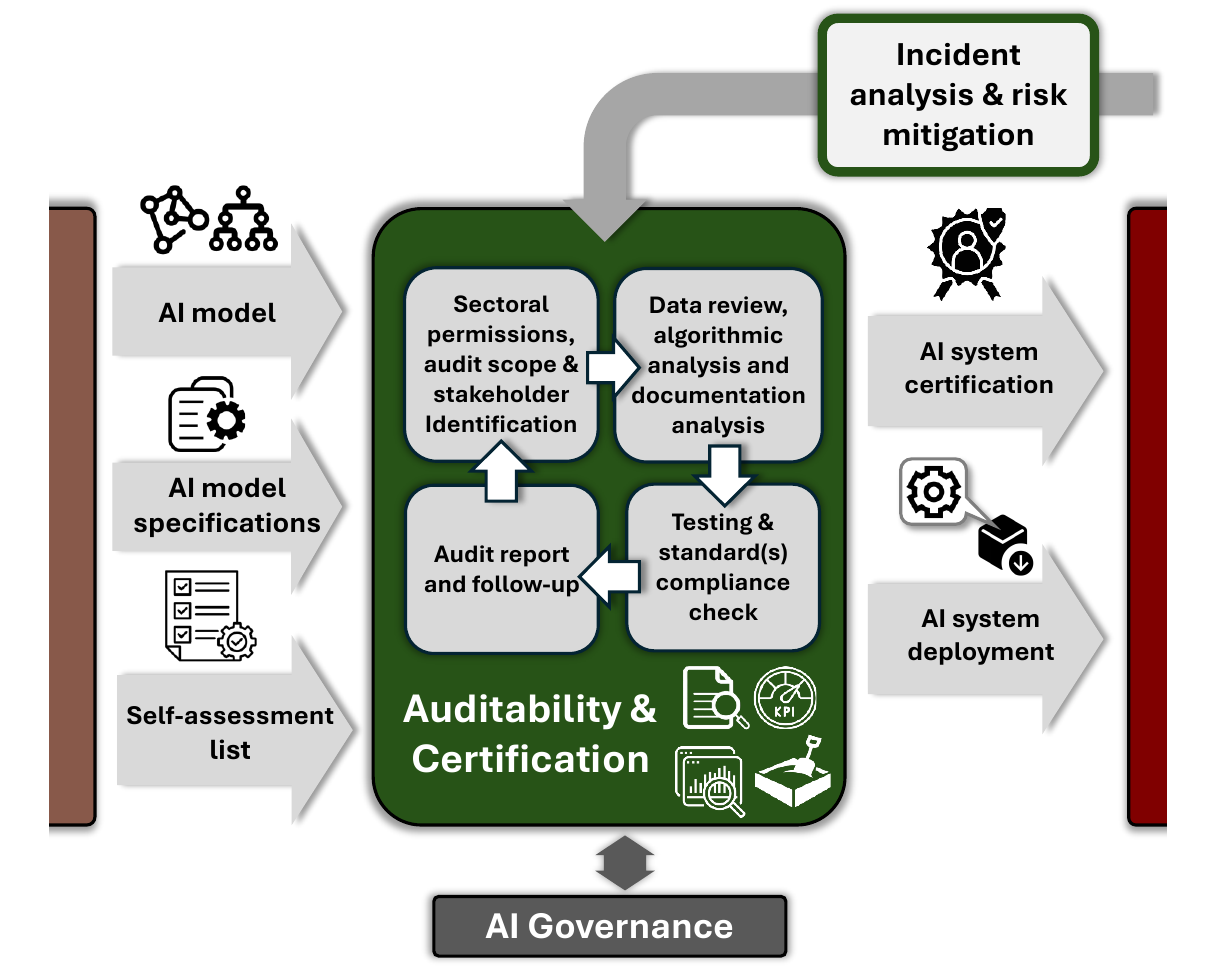}
	\caption{Auditability \& certification within the proposed RAIS framework.}
    \vspace{-3mm}
	\label{fig:Audit}
\end{figure}

A comprehensive discussion of an RAIS in high-risk scenarios necessitates establishing auditability prior to deployment. It is important to first explain why auditability methodologies are required (Subsection~\ref{sec:AUD1}), along with relevant evaluation frameworks and standard norms (Subsection~\ref{sec:AUD-Ass}). Measurement of key attributes, such as robustness, explainability, transparency, traceability, sustainability, and fairness, is also essential. The central question is how to audit an RAIS, which is addressed in Subsection~\ref{sec:AUD-How}. Certification represents the final stage of the auditability process and is the focus of Subsection~\ref{sec:AUD-Cer}. Figure~\ref{fig:Audit} illustrates the auditability and certification dimension within an RAIS framework.

\subsection{Why are Methodologies for Auditability needed?}\label{sec:AUD1}

The process to establish before the design of auditability methodologies requires a dual analysis, which is complementary: 1) the technical TAI requirements mentioned in the previous section; and 2) the legal requirements for the compliance of AI systems with regulations in high-risk scenarios (as a particular case in Europe, those defined in the European AI Act). We refer to \cite{giudici2024artificial} and \cite{fernandez2023trustworthy} for two analyses of TAI requirements made in two different application domains (financial services and autonomous driving, respectively) as a base for auditability.

Auditability is an area that requires significant attention, as it poses major challenges in establishing compliance requirements and metrics tailored to each high-risk scenario. As a previous step to aid accountability, a thorough auditing methodology validates the conformity of the AI-based asset under target to:
\begin{enumerate}[leftmargin=*]
 \item Vertical or sectorial regulatory constraints;
 \item Horizontal or AI-wide regulations (e.g., EU AI Act); and
 \item Specifications and constraints imposed by the application for which it is designed, as ODD features and data quality. 
\end{enumerate}

Auditability refers to the functionality of an AI-based system, which is required, yet not sufficient, to ensure its responsible use. As such, auditability may require transparency (e.g., explainability methods, traceability) or measures to guarantee technical robustness, to mention a few. The auditability of an RAIS may not necessarily cover all requirements for TAI, but rather those foretold by ethics, regulation, specifications, and protocol testing adapted to the application sector (e.g.,  vertical regulation). 

Auditability is a crucial necessity for AI systems, as it can ensure the traceability of decisions and accountability in their operations. Having an auditable RAIS, we can trace and validate decision-making processes, which are essential to identify and correct biases, errors, or unethical practices. This transparency builds trust among users and stakeholders, as they can be confident that the AI systems are working as intended and complying with ethical standards. Moreover, auditability helps maintain compliance with regulatory requirements, which is increasingly important as AI technologies become ubiquitous in almost all sectors. However, extensive auditing introduces trade-offs between compliance and agility. Designing scalable auditing mechanisms that maintain rigor without hindering development speed is therefore essential.

\subsection{Assessments Lists and Standards for Auditability}
\label{sec:AUD-Ass}

To realize an auditable RAIS, it is essential to rely on standardization and assessment lists. These frameworks provide clear guidelines and benchmarks for the development and evaluation of AI systems, ensuring that they meet specific ethical, legal, and technical standards. By following these standardized assessments, organizations can work towards obtaining AI system certifications. 

Therefore, standards and assessments, such as those provided by ISO (International Organization for Standardization), are fundamental for the evaluation of an RAIS. These standards offer a structured framework to ensure that AI technologies adhere to ethical principles, safety protocols, and regulatory requirements. By following established norms, organizations can systematically assess the performance, reliability, and fairness of their AI systems. This not only helps identify and mitigate potential risks, but also fosters transparency and accountability. Implementing these standards ensures that AI systems are developed and deployed in a manner that is trustworthy and aligned with societal values, ultimately improving public confidence in AI technologies.

In 2020, the \textit{"Assessment List for Trustworthy Artificial Intelligence" (ALTAI)} was published \cite{ala2020assessment}. It provides a comprehensive and accessible self-evaluation checklist that guides AI developers and deployments in implementing the principles of TAI in practice. ALTAI translates AI principles into actionable steps, ensuring that AI systems adhere to ethical guidelines, safety protocols, and regulatory requirements.

The use of standardized assessments facilitates interoperability and consistency between different AI systems and applications. This is essential for creating a cohesive and reliable AI ecosystem where systems can work together seamlessly and be evaluated on a common basis. By adhering to recognized standards, organizations can demonstrate their commitment to RAI practices, which can lead to greater trust and acceptance from stakeholders, including users, regulators, and the broader community. Furthermore, audit processes should serve as a dialogue platform among stakeholders, enabling regulators, engineers, and auditors to collaboratively define acceptable trade-offs. Frameworks like ALTAI and Z-Inspection can support this dialogue through structured self-assessment and sector-specific audit workflows. In essence, the integration of assessments and standard norms is a key component in the journey towards building ethical, transparent, and accountable AI systems that benefit society as a whole.

\subsection{How to audit a Responsible AI System?}
\label{sec:AUD-How}

This question is addressed in the context of the AI Act, as it offers a concrete foundation for analyzing current standard requirements within an approved regulatory framework. An RAIS is required to comply with the following seven requirements (AI Act, Chapter 2 \cite{AIA24}), remarking that not all are always required:
\begin{enumerate}[leftmargin=*]
\item Adequate risk assessment and mitigation systems (Art. 9, \emph{Risk management system}).
\item High quality of the datasets that feed the system to minimize risks and discriminatory results (Art. 10, \emph{Data and data governance}; Art. 9, \emph{Risk management system}).
\item Logging of activity to ensure traceability of results (Art. 12, \emph{Record keeping}; 20, \emph{Automatically generated logs}).
\item Detailed documentation providing all the information necessary about the system and its purpose for authorities to assess its compliance (Art. 11, \emph{Technical documentation}; Art. 12, \emph{Record keeping}).
\item Clear and adequate information to the user (Art. 13, \emph{Transparency}).
\item Appropriate human oversight measures to minimize risk (Art. 14, \emph{Human oversight}).
\item High level of robustness, security, and accuracy (Art. 15, \emph{Accuracy, robustness, and cybersecurity}). 
\end{enumerate}

According to this last requirement, auditability must indeed encompass cybersecurity resilience. Certification procedures should incorporate evaluations of domain-specific adversarial robustness, and verification of secure data governance practices. These measures ensure that systems comply with ethical and regulatory requirements while preserving integrity against malicious attacks and hostile environments.

The overarching challenge in auditability and certification is to design metrics and methodologies adapted to the relevant usage scenarios, which are defined in the ODD associated with the AI system and the application at hand. Developing these metrics requires a deep understanding of the AI system's operations and potential risks, ensuring that they are comprehensive and effective.  For time series data, audit protocols should require validation schemes that prevent informational leakage from the validation set into the train set, such as Combinatorial Purged Cross-Validation. These approaches help produce more accurate estimates of the generalization error in situations where labels are computed in overlapping periods, or variables exhibit serial time dependence \cite{de2018advances}.

In \cite{stettinger2024trustworthiness} the authors propose methodologies to ensure trustworthiness for high-risk scenarios in accordance with the EU AI Act. It emphasizes seven key requirements for RAIS to be considered trustworthy and human-centric, integrating concepts like ODD, and introducing Behavior Competency (BC) for risk assessment. The BCs of the automated driving domain are utilized in risk assessment strategies to quantify different types of residual risk. The methodology focuses on a trustworthiness assurance framework, addressing ethical considerations, and includes a roadmap for future AI systems to achieve society's trust and compliance.

\subsection{Certification: The Goal}
\label{sec:AUD-Cer}

The certification of AI systems is the culmination of the auditability process; a necessary step to ensure the safe and effective deployment of AI systems. This process involves a comprehensive evaluation of the design, development and deployment phases of the AI system to verify that it meets established standards and regulatory requirements. Certification provides an official statement that the AI system adheres to ethical principles, safety protocols, and performance benchmarks, instilling confidence in its reliability and trustworthiness.

The importance of this certification process lies in its ability to identify and mitigate potential risks associated with AI systems. By conducting thorough audits, organizations can uncover biases, inconsistencies, and vulnerabilities that can compromise the integrity and effectiveness of the system. This proactive approach helps in addressing issues before they escalate, ensuring that the AI system operates within its intended limits and delivers robust, fair, and transparent results. 

Certification acts as a seal of approval, demonstrating that the AI system has been subjected to meticulous scrutiny and meets the highest standards of quality and safety.  Certification also promotes accountability, as it requires developers and stakeholders to adhere to rigorous standards and best practices throughout the AI lifecycle.

Certification is currently understudied because auditability is the initial step in the development of an RAIS. Currently, standards are being discussed and developed to meet high-level regulatory frameworks established by institutions around the world. For instance, the AI Act has still not come into force, whereas standardization bodies are currently intensively working towards producing the first drafts of the norms and standards to drive auditing processes under such regulations. Consequently, certification must follow later in the process. However, studies related to the certification of different aspects of AI systems have recently originating in the literature \cite{agarwal2023fairness,namiot2024certification}. They expose the growing interest in certification for the deployment and operation of an RAIS.

\vspace{2mm}
\noindent \textbf{\textit{Dimension Frontier.}} The increasing complexity, opacity, and impact of AI systems underscore the urgent need for robust auditability mechanisms. As Li and Goel (2024, 2025) emphasize \cite{li2024making,li2025artificial}, auditability is not merely a technical attribute but a fundamental prerequisite to ensure accountability, transparency, and compliance throughout the AI lifecycle. Their findings highlight the fragmentation and underdevelopment of current audit practices and the lack of detailed guidance to operationalize audits across domains. Very recently, Manheim et al. \cite{manheim2025necessity} further argued that ad hoc or industry-led audits are insufficient and may even be counterproductive, advocating instead the creation of independent AI audit standards boards within safety-critical governance norms, to promote a safety culture similar to that in aviation or pharmaceuticals.

Together, these contributions call for a shift from aspirational ethics to institutionalized oversight, where auditability is proactively embedded in AI system design and where independent, standardized auditing ensures systems evolve responsibly in alignment with societal values.

\section{Accountability and Inspection}
\label{sec:ACC}

\emph{Accountability} establishes liability for decisions derived from the output of an RAIS, once its compliance with the regulations, guidelines, and specifications imposed by the application for which it is designed has been audited. Again, accountability may consist of different levels of compliance with the defined TAI requirements, requiring regular monitoring for its compliance once deployed. 

\begin{figure}[ht]
	\centering
    \includegraphics[width=0.4\columnwidth]{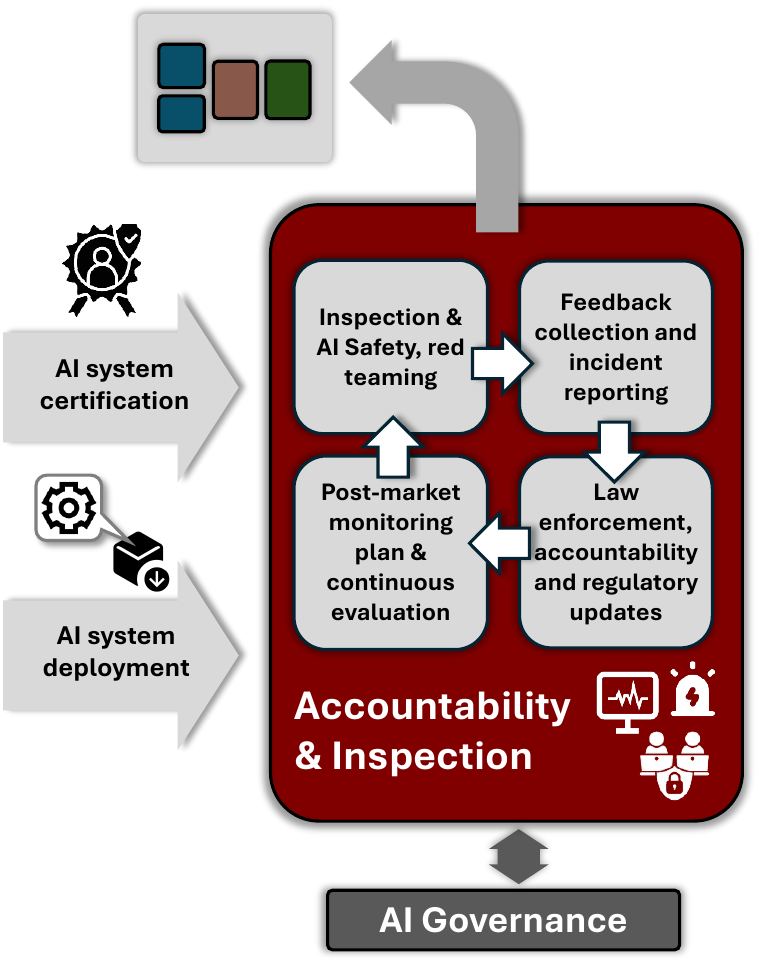}
	\caption{Accountability \& inspection within the proposed RAIS framework.}
	\label{fig:Account}
\end{figure}

Accountability for AI systems is paramount after achieving certification and auditability. Certification ensures that AI systems meet specific standards, but accountability goes beyond initial compliance, comprising several post-market monitoring and inspection mechanisms illustrated in Figure \ref{fig:Account}.

\begin{figure}[ht]
	\centering
	\includegraphics[width=0.4\columnwidth]{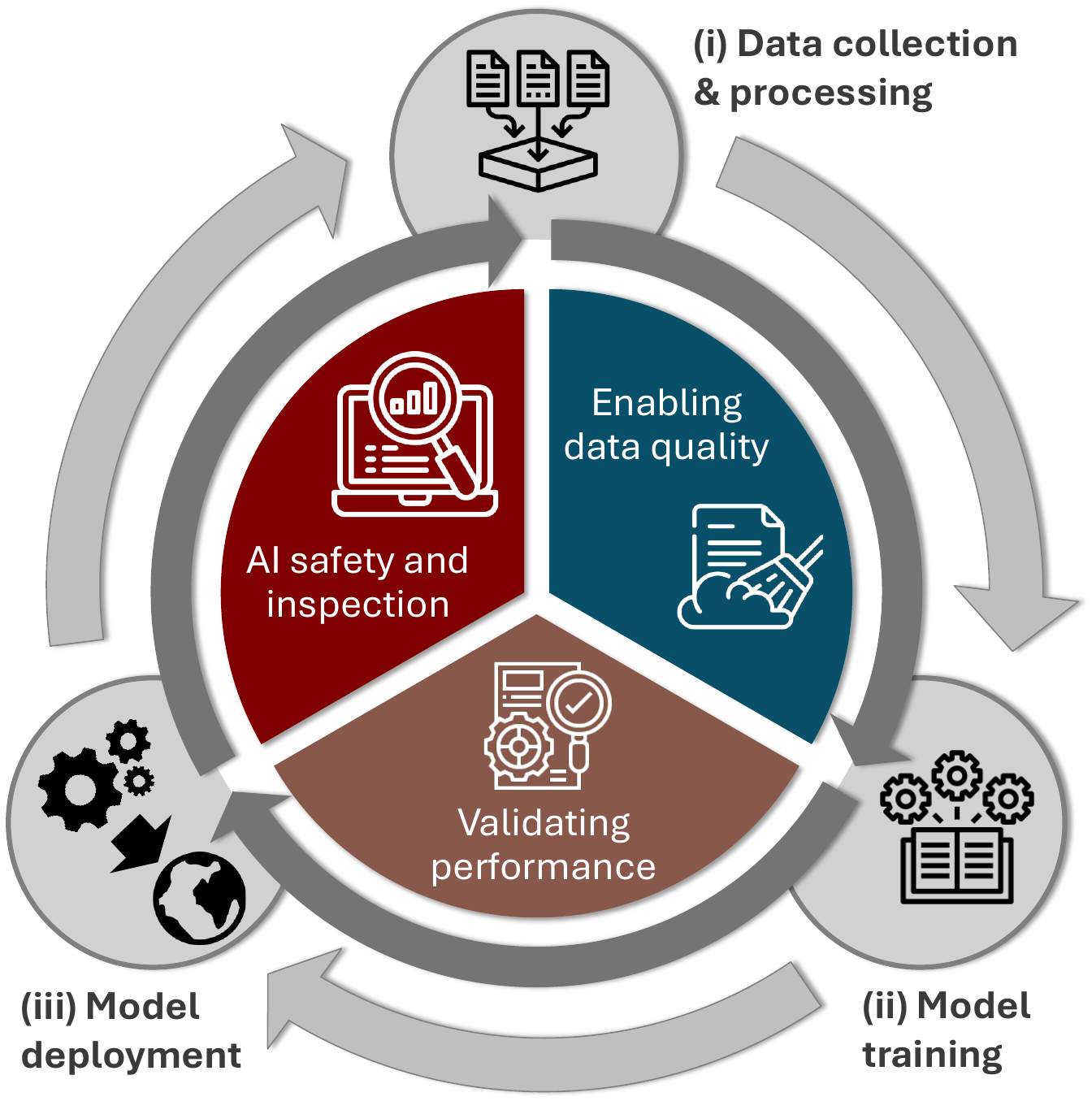}
	\caption{AI lifecycle.}
	\label{fig:fig3}
\end{figure}

Accountability in AI systems is not just a procedural or technical mechanism, but a multifaceted normative concept that integrates ethical responsibility, traceability, and institutional response. As clarified by Novelli et al. (2024)  \cite{novelli2024accountability}, accountability in AI involves the ability to identify the actors responsible for decisions, ensure the resolvability of the results, and enforce the appropriate consequences when harm or failure occurs. This view emphasizes that accountability is not reducible to transparency or liability alone but must include mechanisms for meaningful oversight, justification, and redress. Within the RAIS framework, accountability serves as the cornerstone for aligning AI behavior with societal norms, enabling post-hoc evaluation and reinforcing trust in high-risk environments. By embedding accountability as a structured process, rather than a reactive measure, an RAIS can facilitate institutional trust and ethical compliance throughout the AI lifecycle (Figure \ref{fig:fig3}).

It is important to note the difference between auditability (\emph{ex ante}) and accountability (\emph{post hoc}) when analyzing an RAIS. To achieve continuous post-hoc accountability, it involves continuous monitoring and evaluation to ensure that AI systems operate as intended and adapt to new challenges and contexts. This ongoing accountability is crucial for maintaining trust and reliability in AI technologies.

AI safety is a fundamental goal in this process, ensuring that AI systems are secure, reliable, and resilient to potential risks and threats. By prioritizing accountability and safety, an RAIS can effectively manage potential risks and adapt to new challenges, ensuring their safe and reliable operation. 

\vspace{2mm}
\noindent \textbf{\textit{Dimension Frontier.}} Recent work by Ojewale et al.~\cite{ojewale2025towards} highlights critical gaps in current AI audit tooling and the broader accountability infrastructure. Their analysis identifies persistent fragmentation in the audit ecosystem, limited interoperability between tools, and a lack of clear procedural pathways for integrating auditing within development lifecycles. Importantly, they advocate for a sociotechnical perspective on AI accountability, where tooling is not merely technical, but embedded in organizational workflows, stakeholder engagement, and policy enforcement. This underscores the necessity of building auditability capabilities into systems from the outset while aligning them with external accountability mechanisms. Bridging the accountability–auditability interface is essential for establishing trustworthy AI systems and to support the maturation of AI governance infrastructures.

\section{AI Governance}
\label{sec:AIg}

\begin{figure}[h]
	\centering
    \includegraphics[width=1\columnwidth]{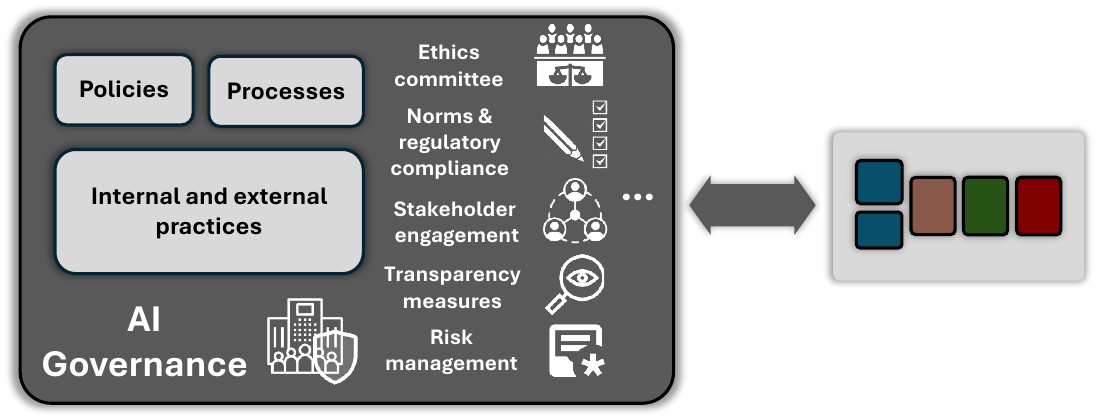}
	\caption{AI governance within the proposed RAIS framework.}
	\label{fig:governance}
\end{figure}

As shown in Figure \ref{fig:governance}, AI governance serves as a core structural element that shapes and is shaped by all other components of the framework. It establishes the institutional and procedural foundations needed to assign responsibilities, ensure regulatory compliance, and continuously oversee system behavior throughout the AI lifecycle.

AI regulation and AI governance are closely related but distinct concepts. Here is a breakdown of their differences:
\begin{itemize}[leftmargin=*]
\item \emph{AI regulation} refers to formal rules, laws, and regulations established by governments or regulatory bodies to oversee the development, deployment, and use of AI technologies. It ensures that AI systems are safe, ethical, and aligned with societal values and protects users and society from potential harm. It focuses on compliance with legal requirements, standards, and policies. 

\item  \emph{AI governance} encompasses the broader framework of policies, practices, and processes that organizations use to responsibly manage AI development and deployment. Ensure that AI systems are developed and used in a way that is ethical, transparent, and accountable within an organization. That is, an RAIS is developed with the goal of helping humanity navigate the adoption and use of AI systems ethically and responsibly. AI governance frameworks include internal and external practices, such as risk management, ethics committees, transparency measures, and stakeholder engagement.
\end{itemize}

Therefore, achieving consensus among states and institutions on AI governance is essential to ensure responsible and ethical development of AI technologies. ensure a society's trust in AI.  An unified approach helps establish consistent standards and regulations, fostering international cooperation and trust. 

There are a few comprehensive proposals for AI governance frameworks. Among them, we highlight the reports on AI governance presented by the United Nations (UN) in 2023 \cite{UN2023} and 2024 \cite{UN2024}, which collected the 2-year efforts of the AI advisory board of this institution. We examine in detail these reports in Subsection \ref{sec:AIg-UN}. A notable proposal is the initial work done by USA, entitled "Framework to advance AI governance and risk management in National Security", which we discuss in Subsection \ref{sec:AIg-USA}. In the private sector, the study on AI governance issues by Google\footnote{Google, "Perspectives on Issues in AI Governance", \url{https://ai.google/static/documents/perspectives-on-issues-in-ai-governance.pdf} [acc. 30/12/2025]} points out five key areas for clarification, explainability standards, fairness appraisal, safety considerations, human-AI collaboration, and liability frameworks.  

\subsection{The Integrating Role of UN AI Advisory Body}
\label{sec:AIg-UN}

From a holistic point of view, AI governance 360º regulates and manages the AI lifecycle, closing the gap that exists between accountability and ethics in technological AI advancement, laying out the path towards the comprehensive development of an RAIS.

\begin{figure}[h!]
	\centering
	\includegraphics[width=0.9\columnwidth]{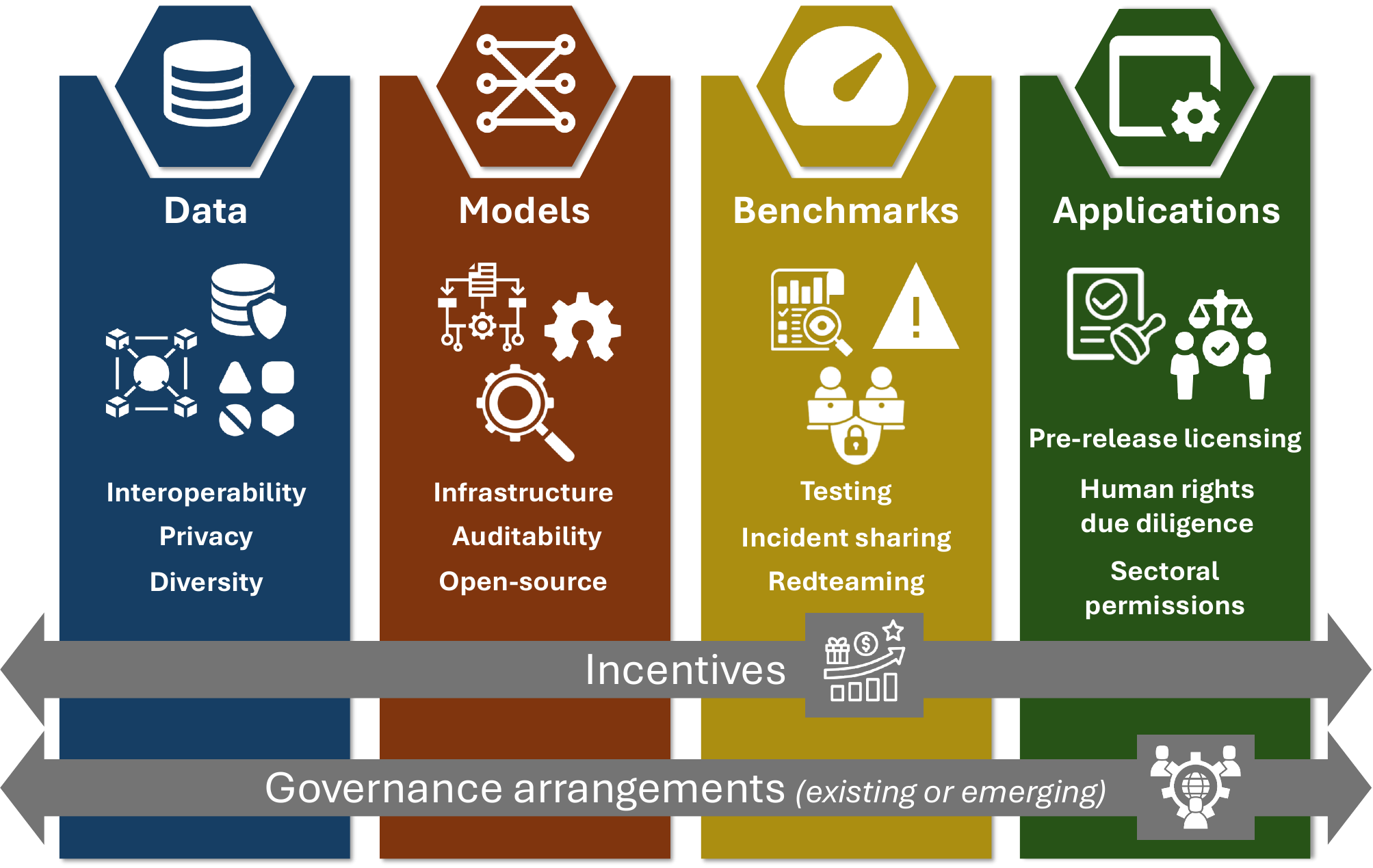}
	\caption{Schema to promote interoperability between different AI governance efforts \cite{UN2023}.}
	\label{fig:fig6}
\end{figure}

Figure \ref{fig:fig6} presents a simplified schema of the AI governance arrangements (both existing and emerging) proposed by the AI Advisory Body. It is intended to promote interoperability between different efforts surrounding AI governance, paying attention to four important aspects: data, models, benchmarks, and the sectoral requirements (e.g. due diligence processes and sector-specific permissions).

During September 2024, the Final Report \textit{``Governing AI for Humanity''} \cite{UN2024} was published. It presents an executive summary reaffirming the imperative of global governance. It highlights the importance of fostering dialogue between countries, acknowledging that although there are differences between nations and sectors, there is a strong commitment to proactive and open communication. Engaging diverse experts, policymakers, business people, researchers and advocates -- across regions, genders, and disciplines -- has shown that diversity should not lead to discordance, and dialogue can set common grounds and foster collaboration effectively. 

UN's Final Report is a valuable document of reflections and conclusions, with 219 points that deserve to be read. 

In parallel to the efforts of the UN AI Advisory Body, on August 9th, 2024 the UN Chief Executive Board for Coordination published the \textit{``White paper on AI governance''}\footnote{UN System White Paper on AI Governance, \url{https://unsceb.org/united-nations-system-white-paper-ai-governance} [acc. 30/12/25]}. The White Paper analyzes the institutional models, functions, and existing international normative frameworks applicable to global AI governance. It is organized into three focus areas, as existing normative and policy instruments, institutional functions and lessons learned from existing governance structures, inclusive normative processes, and agile and anticipatory approaches within the UN system, and presents a set of six general and nine specific recommendations (pages 45 to 47) to further enhance its AI governance efforts. 

The White Paper is a complementary document of the analyzed reports on \textit{``Governing AI for Humanity''}. Together, the three documents present an in-depth analysis of what AI governance should be to enforce a responsible development and use of AI-based systems. 

\subsection{USA Framework to advance AI Governance and Risk Management in National Security}
\label{sec:AIg-USA}

The U.S. \textit{"Framework to advance AI governance and risk management in National Security"}  presents an initial governance framework including four primary pillars:
\begin{enumerate}[leftmargin=*]
    \item The identification of prohibited and high-impact AI use cases based on the risk they pose to national security, international norms, democratic values, human rights, civil rights, civil liberties, privacy, or safety, as well as AI use cases that impact Federal personnel.
      \item The creation of sufficiently robust minimum-risk management practices for those categories of AI that are identified as high impact, including pre-deployment risk assessments.
        \item A catalog and methodologies to monitor the use of high-impact AI systems.
          \item Effective training and accountability mechanisms.
\end{enumerate}

These pillars are developed throughout the 14-page document. Due to the limited space, we do not elaborate further on them. It is a good starting point that we recommend reading and which will need to be expanded to fully address these pillars in practical governance frameworks.

We must point out the fact that  U.S. President Donald Trump (20 January 2025) reversed the executive order passed by former President Joe Biden on October 30, 2023 that aimed to monitor and regulate AI risks (Executive Order 14110) in ,. \footnote{Executive Order 14110 Safe, Secure, and Trustworthy Development and Use of Artificial Intelligence, \url{https://www.govinfo.gov/content/pkg/FR-2023-11-01/pdf/2023-24283.pdf} [acc. 30/12/25]}. The final definition of the US position in terms of regulation is an unknown quantity.

\subsection{Stakeholder Alignment and Conflict Resolution in Responsible AI Governance}
\label{ssec:multistakeholder}

In general, RAIS operate within multi-stakeholder ecosystems where priorities often diverge. Regulators emphasize compliance, risk mitigation, and transparency, while engineers prioritize technical performance, scalability, and innovation. End-users seek usability and fairness, whereas auditors focus on traceability and accountability. These differences, summarized in Table \ref{tab:tensions}, can lead to tensions, such as trade-offs between interpretability and accuracy or between regulatory rigidity and agile development.

\begin{table}[ht]
\centering
\setlength{\tabcolsep}{1pt}

\caption{Stakeholder priorities, conflicts, and resolution methods.}
\label{tab:tensions}
\begin{tabular}{cC{2cm}C{3cm}C{3cm}}
\toprule
\textbf{Stakeholder} & \textbf{Priority} & \textbf{Potential Conflict} & \textbf{Resolution Method} \\
\midrule
Regulators & Compliance, safety & Limits on innovation & Participatory governance, ISO/IEC 42001 \\
\midrule
Engineers & Performance, scalability & Reduced interpretability & XAI, HITL mechanisms \\
\midrule
Auditors & Traceability, accountability & Increased cost/time & Automated audit tools, Z-Inspection \\
\midrule
End-users & Fairness, usability & Complexity of explanations & Role-sensitive XAI, user-centric design \\
\bottomrule
\end{tabular}
\end{table}

To reconcile these conflicts, we advocate for participatory governance models that embed stakeholder engagement throughout the AI lifecycle.  Embedding these practices within iterative feedback loops ensures that stakeholder needs are continuously balanced, promoting trust and societal acceptance of AI systems.

\vspace{2mm}
\noindent \textbf{\textit{Dimension Frontier.}} The challenge of governing AI extends beyond national jurisdictions and demands a coordinated international response. As Roberts et al.~\cite{roberts2024global} emphasize, global AI governance is currently hindered by institutional fragmentation, geopolitical competition, and normative divergence. Rather than relying on the creation of a singular global authority, the authors advocate for reinforcing a flexible and inclusive \textit{regime complex}, namely, a decentralized network of international bodies, norms, and agreements. This approach allows incremental, yet effective, governance while accommodating diverse stakeholder interests, thereby offering a realistic and politically viable pathway forward in AI global oversight.

\section{Illustrative Application of the Proposed Framework: Autonomous Vehicles} \label{sec:example}

To show the practical applicability of the proposed framework, we present a conceptual walkthrough for the high-risk domain of autonomous vehicles, complemented by structured hypothetical scenarios. Autonomous driving systems are explicitly classified as high-risk under the EU AI Act due to their potential impact on safety and fundamental rights. Applying the RAIS framework to this domain illustrates how its five dimensions interact throughout the AI lifecycle:
\begin{itemize}[leftmargin=*]
\item Domain definition: The ODD for autonomous vehicles includes environmental conditions (urban traffic, highways), weather variability, and regulatory constraints. Ethical and safety boundaries are established to prevent harm in edge cases such as pedestrian crossings or emergency maneuvers. Data quality requirements are defined to ensure completeness and representativeness across diverse driving contexts. Standards such as ISO 26262 (\emph{Functional Safety for Road Vehicles}) and ISO/PAS 21448 (\emph{Safety of the Intended Functionality}) guide the definition of safety requirements and hazard analysis.
\item Trustworthy AI design: Technical robustness and safety are prioritized in perception and decision-making modules. XAI techniques are integrated to provide interpretable outputs for critical decisions, such as obstacle avoidance or speed adjustments. Fairness considerations address potential biases in training data, for example, ensuring accurate detection of pedestrians across demographic variations.
\item Auditability and certification: Before deployment, the system undergoes conformity assessments aligned with regulatory standards. Sandbox testing environments simulate high-risk scenarios, such as sudden lane changes or sensor failures, to validate compliance with safety and transparency requirements. Documentation and traceability mechanisms ensure that all design choices and risk mitigation strategies are auditable.

\item Accountability and inspection: Post-market monitoring plans include continuous performance evaluation and incident reporting. For example, if an accident occurs due to sensor malfunction, the accountability process triggers incident analysis, risk mitigation, and system redesign. These steps are followed by re-certification to maintain compliance and trustworthiness.

\item Governance: Governance mechanisms define roles for manufacturers, regulators, and auditors, ensuring participatory oversight throughout the lifecycle. Stakeholder engagement is embedded in feedback loops, enabling adaptive responses to emerging risks and regulatory updates.
\end{itemize}

To further illustrate these principles, we propose structured hypothetical scenarios:
\begin{itemize}[leftmargin=*]
\item Scenario 1: A pedestrian suddenly crosses in poor visibility. The system’s decision-making process is explained through XAI, enabling auditors to verify compliance with safety protocols and ethical guidelines.
\item Scenario 2: Bias is detected in traffic sign recognition due to under-representation of rural signage. Auditability mechanisms identify the issue, triggering corrective actions such as dataset enrichment and model retraining.
\item Scenario 3: A post-deployment collision prompts accountability procedures, including incident analysis, ODD adjustment, and re-certification under updated safety standards.
\end{itemize}

These examples illustrate how the proposed framework can be applied in practice, showing its ability to connect technical design, regulatory compliance, and governance in a coherent way. By walking through a high-risk domain such as autonomous vehicles and exploring hypothetical scenarios, we demonstrate that the framework is not only conceptual but also adaptable to real-world scenarios.

{\section{Design Insights, Reflections and Challenges}
\label{sec:Key}

This section consolidates conceptual insights derived from the proposed framework and the synthesis of existing literature together with a reflection on challenges.

Subsection \ref{sec:Design} aims to highlight the most relevant design considerations to operationalize an.  Subsection \ref{sec:Challenge}  outlines a reflective discussion of the challenges  that must be addressed to ensure that these principles can be translated into actionable workflows in diverse high-risk domains.}

\subsection{Design Insights} \label{sec:Design}

In what follows, we outline crucial considerations for design insights to align the RAIS framework with social values and global governance trends, ranging from TAI and certification practices to inclusivity, innovation, and regulatory agility.
\begin{enumerate}[leftmargin=*]

\item TAI is a critical paradigm for meeting upcoming regulations, addressing ethical issues, and managing risk analysis in human-AI collaboration and interaction, ensuring AI governance and technical soundness of an RAIS.

\item Certified AI systems are more likely to be trusted and accepted by users, regulators, and the broader community, facilitating their adoption and integration into critical applications. Certification enhances the credibility of AI systems and promotes their responsible and ethical use.

\item Aligned with the AI Seoul Summit \cite{Seoul2024}, held in May 2024, the summit emphasized the importance of AI governance discussions to promote safety, innovation, and inclusion, 
with the aim of shaping a global strategy for responsible AI development.  

\item The development of the RAIS framework lies in the convergence of regulation and innovation. By balancing these two aspects, we can boost innovation and create an RAIS that is not only powerful and efficient, but also trustworthy and accountable. This approach will help build societal trust in AI, ensuring that its benefits are realized while minimizing potential risks and harms.

\item The rapid advancement of AI systems, driven by significant investments and continuous research, underscores the need for agile methodologies throughout the RAIS framework, such as the ones proposed in this work. Emerging AI discoveries can introduce new threats, misuses, and challenges in real-world applications. Therefore, agile regulatory frameworks, standards, and norms are essential to enable the development of auditability, accountability, and AI governance methodologies that keep pace with the evolving landscape of AI.

\item Multi-stakeholder alignment is not optional, but foundational: conflict resolution requires co-design practices, lifecycle oversight, and transparent communication supported by explainability and governance standards.
\end{enumerate}

\subsection{Challenges}
\label{sec:Challenge}

Although promising, the development of an RAIS also presents a complex landscape of reflective challenges to ensure their safe, ethical, and accountable deployment. These challenges reveal gaps in current practice, but also reveal opportunities for interdisciplinary collaboration and regulatory innovation to align AI progress with societal expectations:
\begin{enumerate}[leftmargin=*]

\item We first draw attention to the need for auditability metrics for RAIS analysis and compliance. Adherence to established future standardizations and guidelines (such as those provided by ISO and/or IEEE standards) ensures that an RAIS is developed and deployed responsibly.

\item Explainability plays an essential role in this scenario. Communication with stakeholders and explanations helps build trust and ensure that AI systems meet societal expectations. 

\item Human-AI collaboration is expected to shape the future. Although human decision-making remains crucial in high-risk scenarios, effective integration, trust, and collaboration with AI are essential to improve productivity and human augmentation. 

\item The stakeholder context leads us to a fundamental role for XAI \cite{bello2025three} and human-AI collaboration. Concrete strategies for stakeholder participation, particularly in participatory governance, co-design practices, and lifecycle oversight, must be developed  to enhance the applicability of the framework.

\item Safety and security play an essential role in all AI systems, especially those that interact in an open world. Continuous monitoring and evaluation of AI systems are necessary to identify and address emerging issues. As an example, let us turn our attention to a context of real-life real application with continuous safety and security challenges (e.g., automated vehicles \cite{ullrich2025ai}).

\item As highlighted in \cite{bengio2024managing}, it is necessary to have a comprehensive plan combining technical research and adaptive governance to effectively address challenges, to achieve consensus to manage extreme AI risks. 

\item Agentic AI systems based on LLMs introduce a new frontier of challenges for responsible AI. These systems are capable of generating content and making autonomous decisions, adapting goals, and interacting with dynamic environments. This evolution demands continuous oversight, dynamic value alignment, and new safety mechanisms beyond static predeployment checks. As highlighted by Schneider \cite{schneider2025generative}, trustworthiness in agentic LLMs requires rethinking traditional audit and accountability to include real-time monitoring and fail-safe design patterns.

\item A final challenge lies in translating the proposed framework into actionable workflows across diverse sectors. While the current version of the framework provides a reference model, its practical adoption requires domain-specific instantiations that account for sectoral regulations, operational constraints, and context-sensitive risk profiles. For example, sectors such as healthcare, finance, and autonomous mobility impose distinct ethical, technical, and compliance requirements that demand a tailored RAIS framework rather than general and uniform approaches. Developing tailored methodologies, metrics, and governance mechanisms for these sectors will demand interdisciplinary collaboration and iterative refinement. Future research should focus on creating modular extensions and operational guidelines that enable organizations to adapt the framework effectively, ensuring both technical feasibility and regulatory alignment.
\end{enumerate}

\vspace{2mm}
\noindent \textbf{\textit{Responsibility Statement.}} These reflections underscore the complexity and importance of developing an RAIS that prioritizes safety, ethics, and accountability. They serve as both a reflection of the current state of the field and a guide for future efforts, highlighting the need for interdisciplinary collaboration, agile regulatory frameworks, and alignment of technological advances with societal values.

\section{Conclusions} \label{sec:Con}

This work has presented a comprehensive RAIS framework that responds to the three research questions posed in the Introduction. First, the framework operationalizes auditability and accountability through a lifecycle-based architecture that connects design, certification, and post-deployment monitoring, enabling both proactive and reactive compliance. Second, it introduces a methodology that integrates technical trustworthiness with socio-legal requirements by unifying TAI principles with governance mechanisms, offering practitioners a structured pathway to meet regulatory obligations while maintaining system robustness. Third, it embeds dynamic feedback loops and participatory governance to ensure continuous alignment with societal values and evolving risks.

 For practitioners, this framework provides a practical blueprint for developing an RAIS in high-risk domains, bridging technical design and institutional accountability through actionable steps such as self-assessment lists, conformity checks, and governance practices.

Our paper also calls for collective action, laying the foundation for a new generation of RAISs that are robust, accountable, and attuned to societal values. As the AI landscape continues to evolve, the RAIS framework offers both a compass and a toolkit: supporting practitioners, policymakers, and researchers in advancing complex AI-based systems that are not only powerful, but also principled and accountable. 

\bibliographystyle{ieeetr}
\bibliography{2-references}

\end{document}